\documentclass[12pt, draftclsnofoot, onecolumn,journal]{IEEEtran}

\usepackage{amsmath,amsfonts,amssymb,amsbsy,url,verbatim}
\usepackage{times}
\usepackage[english]{babel}
\usepackage[dvips]{graphicx}
  \usepackage{subfigure}
\usepackage{bm}
\usepackage{comment}
\usepackage{stfloats}
\usepackage{cite}
\usepackage{psfrag}
\usepackage{dsfont} 
\usepackage{wrapfig}
\usepackage{algorithm}
\usepackage{algcompatible}

\newcommand{\diag}{\mathrm{diag}}

\newcommand{\back}{\!\!\!\!\!}

\newcommand{\wmpg}{0.485\linewidth}

\graphicspath{%
     {./figs/}%
}

\begin{document}

\renewcommand{\tablename}{\textsc{TABLE}}

\title{Millimeter Wave Receiver Efficiency:\\ A Comprehensive Comparison of Beamforming Schemes with Low Resolution ADCs
}
\vspace{-1mm}

\author{%
    \IEEEauthorblockN{Waqas~bin~Abbas, Felipe~Gomez-Cuba, Michele Zorzi\vspace{0mm}}\\
    \IEEEauthorblockA{\small Department of Information Engineering University of Padova, Italy \vspace{-0.5mm}}\\
    \IEEEauthorblockA{\small E-mail: \texttt{\small \{waqas,zorzi\}@dei.unipd.it}, }\texttt{\small fgomez@gti.uvigo.es}%
}

\maketitle

\begin{abstract}
In this work, we study the capacity and energy efficiency of Analog, Hybrid and Digital Combining (AC, HC and DC)  for millimeter wave (mmW) receivers. Our comparison extends previous work by taking into account the power consumption of all components of the receiver, not just Analog-Digital Converters (ADC), by considering the practical limitations of beamforming in each architecture, and by developing a performance chart representation that enables comparison of different architectures at a glance.
We consider an Additive Quantization Noise Model (AQNM) where the achievable rate is limited both by the unquantized SNR and by the number of ADC bits. 
We show that the usual claim that DC requires the highest power stems from a quite conservative focus only on ADCs. Rather, considering the power consumed by all mmW receiver components, the margin by which HC outperforms DC is narrow and very fragile, and easily reverted by state-of-the-art improvements in ADCs, or when there is any mismatch between the number of RF chains built in the HC and the actual rank of the channel.
Our analysis shows that AC is only preferable if the channel rank is strictly one or the power constraint is very stringent. HC, in turn, is only preferable if the ADC technology is inefficient (i.e., when considering ten year old devices) \textit{and} the channel rank is very small (but greater than one). In any other situation, DC is preferable, and due to its versatility and the uncertainty on future mmW conditions, it is the most future-proof architecture.
\end{abstract}

\begin{IEEEkeywords}
\noindent Millimeter Wave, Analog Beamforming, Hybrid Beamforming, Digital Beamforming, Energy Efficiency, Spectral Efficiency, Low Resolution ADCs
\end{IEEEkeywords}

\thanks{\scriptsize Part of this work was presented at the European Wireless 2016 conference, Oulu, Finland. This project has received funding from the European Union's Horizon 2020 research and innovation programme under the Marie Sk\l{}odowska-Curie grant agreement No 704837.}


\section{Introduction}
\label{sec:intro}
The millimeter wave (mmW) spectrum (30-300 GHz), where a very large bandwidth is available, is considered as a prime candidate to fulfill the data rate requirements of future broadband communication \cite{5GWillWork,Andrews5Gbe,Boccardi5DT5G,KhanFmmWave,Rap2014mmPotCha}.
However, communication at these frequency bands exhibits high pathloss \cite{5GWillWork}.
To overcome this high pathloss, spatial beamforming/combining using large antenna arrays is considered as an essential part of a mmW communication system. 

Analog, Hybrid and Digital are the available beamforming/combining schemes for MIMO communication systems \cite{SMBF2014}.
Although a fully digital architecture, which requires a separate RF chain per antenna element, is a popular choice in classical systems, it is not generally considered as a viable option for large antenna arrays due to the high power consumption of analog-digital signal processing components \cite{mmWBF2014}.
This is even more critical at mmW frequencies where the power consumption of an Analog to Digital Converter (ADC) in a receiver grows linearly with the system bandwidth. 
However, multiple options exist to reduce the power consumption
\begin{enumerate}
 \item The use of Analog Combining (AC), which requires a single Radio-Frequency (RF) plus ADC chain, consumes least power and is an attractive choice whenever the advantages of digital processing techniques (mainly spatial multiplexing) are not required \cite{AlkhateebMIMOSolMag}. 
 \item The use of Hybrid Combining (HC), which performs combining in both the analog and the digital domain with a reduced number of RF chains, at the cost of lower flexibility than fully digital architectures \cite{AyachHC}.
 \item The use of low-resolution ADCs (for example, 1 bit) while maintaining the Digital Combining (DC) MIMO architecture \cite{MoHeath1bit,MassMIMO1bit}.
\end{enumerate}

In our recent work \cite{MyPCCompEW16}, it is shown that the popular perception that DC has the worst power efficiency is not always accurate, especially when taking into account the power consumption of all the receiver components. Rather, there are regimes where DC can still result in a lower power consumption than the hybrid combining (HC) architecture. 

In this paper, we extend our analysis and provide a comprehensive characterization of the Spectral Efficiency (SE, defined as capacity/bandwidth) versus the Energy Efficiency (EE, defined as capacity/total power consumption) of AC, HC and DC architectures for a wide variety of possible scenarios. Our results show that in a majority of use-cases DC is a more attractive choice than HC in terms of both SE and EE. Primarily, we argue that the popular notion that DC has the highest power consumption, based solely on ADC power, is quite narrow. Rather, DC with low resolution ADCs may result in a lower power consumption than HC when the power consumption of all receiver components is taken into consideration. In our analysis we show that 

\begin{itemize}
\item The achievable rate of a quantized received signal is directly related to the number of ADC bits. As SNR increases, the minimum number of bits required to achieve nearly the same SE than the unquantized system also increases.
\item In all combining schemes, there is an optimal number of bits that maximizes the EE. This optimal quantization resolution is directly related to the SNR and is inversely related to the ADC power figure of merit.
\item The AC scheme is only the best option in terms of both EE and SE when a mmW channel consists of only 1 propagation path and/or in very low SNR regimes, and/or under a very stringent power constraint. 
\item In many cases DC outperforms HC in terms of both SE and EE. There are a few use-cases where HC can be a preferable option than DC, but this occurs only when all the system properties are stacked in favor of HC at the same time: the channel has a very low spatial multiplexing advantage (i.e., with only few independent paths), the receiver has a very low number of RF chains to number of antennas ratio, the number of RF chains matches very well the degrees of freedom of the channel, and a very conservative pessimistic high power ADC model is considered.
\end{itemize}

In summary, we show that when considering the total power consumption of the receiver instead of just focusing on the ADC power consumption, in a majority of scenarios DC may be a preferable option than HC. Also, due to the fact that future mmW system characteristics are not yet completely settled, DC appears to be the most future-proof technology. AC and HC may be better choices in their respective special scenarios, but these conditions are the exception, not the norm.

\subsection{Related Work}
\label{ssec:Rel_Work} 

Recent works study energy efficient designs, particularly focusing on how the system capacity varies as a function of the ADC resolution. In \cite{confiwcmcNossekI06}, the capacity of a quantized MIMO system along with coding has been studied, while in \cite{MurrayAGCQuant} the capacity and bit error rate for a quantized MIMO system was analyzed. In \cite{AmineGaussQaun} a capacity lower bound for a quantized MIMO system with Gaussian input were analyzed.
However, the quantized capacity model is only optimal at low SNR regimes.
In \cite{Madhow_1bitADC}, an exact nonlinear quantizer model is utilized to evaluate the optimal capacity for a 1-bit ADC. In \cite{OptADCRes}, considering a MIMO channel and an additive quantization noise model (AQNM, an approximate model for ADCs), a joint optimization of ADC resolution and number of antennas is studied.
In a recent work \cite{OrhanER15PowerCons}, the authors studied how the number of ADC bits $b$ and the bandwidth (sampling rate) $B$ of ADCs affect the total power consumption for AC and DC based receivers with a stringent power constraint such as a mobile station. They studied the optimal $b$ and $B$ which maximize the capacity for AC and DC only for low power receiver design while also showing that DC with similar power budget to AC may achieve a higher rate than AC when the channel state information is available at the transmitter. 
In another recent work \cite{Ahmed16_EE}, the EE and SE for low resolution ADC HC architecture is studied. The authors only show the advantages of a low resolution HC with few RF chains over an infinite resolution ADC for DC and HC.
 
Recently, energy efficient architectures for HC are proposed \cite{GaoHeathHC,RialandHeath}.
In \cite{GaoHeathHC}, an energy efficient HC architecture has been proposed where each RF chain is only connected to a subset of antennas.
In \cite{RialandHeath}, to further reduce the power consumption of a conventional HC with phase shifters, switch based architectures are proposed, where at a particular instant only a reduced set of antennas (equal to the number of RF chains) is selected and connected to the RF chains. 
However, in both works the proposed architectures result in a lower SE than the fully connected phase shifter architecture.

In \cite{FanULRateLowADC15,ZhangSELowADC}, the SE of uplink massive MIMO with low resolution ADCs is studied, and it is shown that few ADC bits are enough to achieve almost the same SE of unquantized MIMO, and in \cite{FanULRateLowADC15} it is also shown that a 2-bit ADC achieves good performance for a small Rician k-factor.

To the best of our knowledge, there are no previous works that analyze and compare AC, DC and HC for a finite resolution ADC (i.e., quantized MIMO system), and study their trade-off in terms of both EE and SE. We would like to highlight that our model identifies and overcomes some limitations of previous frameworks on this topic. 
\begin{itemize}                                                                                                                                                                          \item First, some works on HC benchmark their proposal against ideal non-quantized DC; thus, rather than evaluating which architecture requires less power, these works focus on how close the HC rate is to ideal DC. This means that these works have an unverified initial assumption that HC has superior EE and is preferable to a quantized DC with a carefully chosen low number of bits.                                                                                                                                                                                                                                                                                                                                                    \item Second, it is frequent that only the power consumption of ADCs is taken into account, under the implicit assumption that this is completely dominant. However, some brief literature review shows that the power consumption of Phase Shifters (PS) is in the order of tens of mW  \cite{LiPSPow,Rx_Pow_LNA_PS_C},
and thus a HC system with 64 antennas and 4 RF chains can easily burn over 2.5 W already in the analog stage preceding the ADCs. 
\item Third, some works do not include some components in the HC architecture, such as the signal splitters (SP) needed to derive the signal from one antenna to multiple PS arrays. These components, too, have a non-negligible power signature. Our analysis takes into account the power consumption of \textit{all} the components of the receivers as considered in the recent literature \cite{OrhanER15PowerCons,RialandHeath}. 
\item Fourth, the power consumption figure of merit for ADCs varies significantly between different references (4 orders of magnitude). We discuss the origin of this variation, and classify the values in three categories: a decade-old legacy hardware, an existing state-of-the-art device, and a plausible future projection based on an existing hardware survey. It is important to note that HC only outperforms DC with the legacy ADCs, and thus the existing literature based on pessimistic ADC figures of merit has a tendency to overestimate the benefits of HC. 
\end{itemize}
  
In order to facilitate the verification of these four claims by other independent researchers, we have published a web viewing tool\footnote{\vspace{-10mm} \url{https://dl.dropboxusercontent.com/u/1770302/mmWaveADCwebviewer/index.html}} for our simulation. In the application,  the power per component may be modified by the reader, in order to assess the influence of each part of the receiver. Moreover, previous results without taking some components into account may be reproduced by entering zeros in the appropriate fields.
  
\section{System Model}
\label{sec:model}

In 5G mmW cellular systems, a distinction is usually made between the Base Stations (BS) and the User Equipment (UE). Typical values for the UE are $16$ antennas and $1$ W power, while typical values for the BS can be $64$ or more antennas and $5$ W power in small pico-cells, or $50$ W in large macro-cells. Moreover, the system bandwidth varies from $500$ MHz up to $7$ GHz \cite{OrhanER15PowerCons}.

For the analysis in this paper, however, the roles of UE and BS are not that relevant. Instead, we simply distinguish between transmitter and receiver roles (that may be taken by either the BS or the UE), and we obtain analytical expressions for SE and EE as a direct function of the number of antennas and of the SNR of the link, obtaining results that apply to all mmW networks with these two values.
Note that we considered finite bits only at the receiver while the transmitter is equipped with a full precision Digital-to-Analog converter.

We consider a point-to-point multiple input multiple output (MIMO) mmW channel where the transmitter is equipped with $N_t$ transmit antennas and the receiver with $N_r$ receive antennas.  
The channel has bandwidth $B$ and a delay spread much smaller than the symbol period $D_s \ll 1/B$, so there is no inter-symbol interference and the received signal can then be given as \cite{rappaport2014millimeter}

\begin{equation}
\textbf{y} = \textbf{H}\textbf{x} + \textbf{n}
\label{eq:y}
\end{equation}
where $\textbf{x}$ and $\textbf{y}$ represent the transmitted and the received symbol vectors at discrete time instants with period $1/B$, respectively, $\textbf{n}$ is the i.i.d circularly symmetric complex Gaussian noise vector, $\textbf{n} \sim \mathcal{CN}(\textbf{0},N_o\textbf{I})$, and $\textbf{H}$ represents the $N_r \times N_t$ channel matrix that varies following a fast block-fading model that remains constant for a small number of symbols and takes independent identically distributed values across blocks; this means that the capacity of the system is the \textit{ergodic capacity} (average mutual information over the realizations of $\textbf{H}$).
The mmW channel matrix is randomly distributed following a random geometry with a small number of propagation paths (order of tens) grouped in very few clusters of similar paths (average $1.9$)\cite{KhanFmmWave,Alkhateeb14_ChanEst} 

\begin{equation}
   \textbf{H} = \sqrt{\dfrac{N_{t}N_{r}}{\rho N_cN_p}}\sum_{k=1}^{N_c}\sum_{\ell=1}^{N_p}g_{k,\ell}\textbf{a}_{r}(\phi_{k}+\Delta\phi_{k,\ell}) \textbf{a}_{t}^H(\theta_k+\Delta\theta_{k,\ell})
\end{equation}
 
where $\rho$ is the distance dependent path-loss, $N_c$ is the number of independent clusters, $N_p$ represents the number of paths per cluster, $g_{k,\ell}\sim \mathcal{CN}(0,1)$ is the small scale fading associated with the $\ell^{th}$ path of the $k^{th}$ cluster, $\phi_k$ and $\theta_k$ $\in [0,2\pi)$ represent the mean angle of arrival (AoA) and angle of departure (AoD) of the $k^{th}$ cluster at the receiver and at the transmitter, respectively. The AoA and AoD of each path within each cluster vary around the mean direction of that cluster, with a standard deviation $\theta_{RMS}$. We represent by $\Delta\phi_{k,\ell}$ and $\Delta\theta_{k,\ell}$ $\sim \mathcal{N}(0,\theta_{RMS}^2)$ the differential AoA and AoD of the $\ell^{th}$ path of the $k^{th}$ cluster.

Here, we model the antenna arrays at both the transmitter and the receiver as uniform linear arrays (ULA) with adjacent antenna spacing of half the wavelength of the transmitted signal ($\lambda/2$). Under this model, a spatial signature vector $\textbf{a}_{t}$ for the transmit array can be expressed as a function of the AoD as follows
\begin{equation}
   \textbf{a}_{t} = \dfrac{1}{\sqrt{N_{t}}}[1, e^{j{\pi}\sin (\theta)}, ... , e^{j(N_{t} - 1){\pi}\sin (\theta)}]^T 
\end{equation}
where $T$ represents the transpose.
Likewise, an analogous expression characterizes the spatial signature vector for the receiver, $\textbf{a}_{r}$.
Finally, for a 28 GHz channel, the path-loss is computed with \cite{Akdeniz_mmW_CM}
$$\rho_{LOS} (dB)=61.5+20\log_{10} (d)+\xi,\; \xi\sim\mathcal{N}(0,5.8),$$
$$\rho_{NLOS} (dB)=72+29.2\log_{10} (d)+\xi,\; \xi\sim\mathcal{N}(0,8.7),$$
where $d$ represents the distance between the transmitter and the receiver on a straight line while the variation of distances traversed by different paths is captured in $g_{k,\ell}$.

The parameters suggested to model mmW 28 GHz channels in the literature \cite{Akdeniz_mmW_CM} are $N_p=20$, $N_c\sim\max\{\textnormal{Poisson}(1.8),1\}$ and $\theta_{RMS}\sim 10^o$. In our simulations, however, we set specific non-random values for $N_c$ to study the effect of the rank of the channel matrix $\mathbf{H}$ in the performance of receiver architectures. We consider a rank-1 channel with $N_c=N_p=1$; a low-rank channel with $N_c=1$ and $N_p=10$, where $\mathbf{H}$ typically has $4-5$ dominant eigenvalues; and a not-so-low rank channel $N_c=2$, $N_p=10$ where $\mathbf{H}$ typically has $8-10$ dominant eigenvalues. 

\subsection{Quantized Received signal}
\label{ssec:Quan_Rx}
The capacity of the quantized MIMO channel with a 1 bit ADC under an exact non linear quantization model is shown in \cite{Madhow_1bitADC}.  
However, it is difficult to extend this exact non-linear model to a higher number of bits. A common lower-bound approximation to the capacity of quantized systems consists in modeling the quantization as an additive Gaussian noise with power inversely proportional to the resolution of the quantizer, that is, $2^{-b}$ times the receiver input power where $b$ is the number of ADC bits. In recent studies \cite{OrhanER15PowerCons}, \cite{FanULRateLowADC15}, this Additive Quantization Noise Model (AQNM) has been applied to the study of mmW quantized signal modeling with an arbitrary number of ADC bits.

\subsection{Received Signal Model with AQNM}
\label{ssec:AQNM_Rx}

We consider that the received signals at each antenna may be subject to some analog processing prior to quantization. This RF-processed received signal is converted to the digital domain by multiple ADCs (one ADC for each inphase and quadrature component for each vector dimension). In the AQNM \cite{AQNM07}, we represent the quantized version $\textbf{y}_q$ of the received signal (\ref{eq:y}) as 

\begin{equation}
\textbf{y}_q = (1-\eta)(\textbf{H}\textbf{x} + \textbf{n}) + \textbf{n}_q
\label{eq:yq}
\end{equation}

where $\textbf{n}_q$ is the additive quantization noise and $\eta$ is the inverse of the signal-to-quantization noise ratio, which is inversely proportional to the square of the resolution of an ADC (i.e., $\eta \propto 2^{-2b}$).
For a Gaussian input distribution, the values of $\eta$ for $b \leq 5$ are listed in Table \ref{tab:etavsb}, and for $b > 5$ can be approximated by  $\eta = \frac{\pi \sqrt{3}}{2} 2^{-2b}$ \cite{FanULRateLowADC15}. 
We denote by $\gamma_q$ the signal-to-noise ratio SNR of $\textbf{y}_q$, given by

  \begin{table}
     \centering
     \caption{$\eta$ for different values of $b$ \cite{FanULRateLowADC15}}
     \begin{tabular}{|c||c|c|c|c|c|}
         \hline
         $b$ &  1 & 2 & 3 & 4 & 5\\ 
         \hline
         $\eta$  & 0.3634 & 0.1175 & 0.03454 & 0.009497 & 0.002499 \\
         \hline
     \end{tabular}
     \label{tab:etavsb}
 \end{table}

\begin{equation}
	\gamma_q = \dfrac{(1-\eta)^2(\textbf{H}\textbf{R}_{\textbf{xx}}\textbf{H}^H)}{(1-\eta)^2{N_o}\textbf{I} + \textbf{R}_{\textbf{n}_q\textbf{n}_q}}
	\label{eq:gammaQ}
\end{equation}

where $\textbf{H}\textbf{R}_{\textbf{xx}}\textbf{H}^H$ is the received signal power at the output of the quantizer, $\textbf{R}_{\textbf{xx}}$ is the input covariance, $N_o$ is the noise power, and  $\textbf{R}_{\textbf{n}_q\textbf{n}_q} = \eta(1-\eta)(\textbf{H}\textbf{R}_{\textbf{xx}}\textbf{H}^H) + N_o\textbf{I}$ \cite{AQNM07} is the covariance of the quantization noise. 

Substituting $\textbf{R}_{\textbf{n}_q\textbf{n}_q}$ in (\ref{eq:gammaQ}) yields
 \begin{equation}
	\gamma_q = \bigg|\dfrac{(1-\eta)(\textbf{H}\textbf{R}_{\textbf{xx}}\textbf{H}^H)}{{N_o}\textbf{I} + \eta(\textbf{H}\textbf{R}_{\textbf{xx}}\textbf{H}^H)}\bigg{|},
	\label{eq:gammaQ_nor}
\end{equation}
%
%
and finally, in terms of the SNR of the unquantized signal ($\gamma$), $\gamma_{q}$ can be written as \cite{Barati_IBF} 
\begin{equation}
 \gamma_{q} = \frac{(1-\eta )\gamma}{1+\eta \gamma}
 \label{Eq:SNReff}
\end{equation} 
At low SNR, $\gamma_q$ can be approximated as $(1 - \eta)\gamma$, while at high SNR and for finite bits $b$, the quantized SNR $\gamma_q$ is tightly upper bounded by $\min(\frac{1-\eta}{\eta},\gamma)$. 
Note that, for a very high resolution, $\eta \rightarrow 0$, and $\gamma_q$ in (\ref{eq:gammaQ_nor}) will be equal to the SNR of the unquantized signal $\gamma$.  
Finally, the capacity of the MIMO link with an AQNM signal in (\ref{eq:yq}) is given as

 \begin{equation}
	C_q = \mathrm{E}_{\mathbf{H}}\left[\max_{\textbf{R}_{\mathbf{xx}}}B\log_2\bigg|\textbf{I} + \dfrac{(1-\eta)(\textbf{H}\textbf{R}_{\textbf{xx}}\textbf{H}^H)}{{BN_o}\textbf{I} + \eta(\textbf{H}\textbf{R}_{\textbf{xx}}\textbf{H}^H)}\bigg|\right]
	\label{eq:Cq_nor}
\end{equation}
\vspace{2mm}

\begin{figure}[t]
    \centering
    \begin{minipage}[t]{\wmpg}
    \psfrag{Nt = 64, Nr = 16, Npath = 8, LPADC}{\back \scriptsize $N_t = 64$, $N_r = 16$, $N_{paths} = 8$, LPADC}
        \includegraphics[width=\columnwidth]{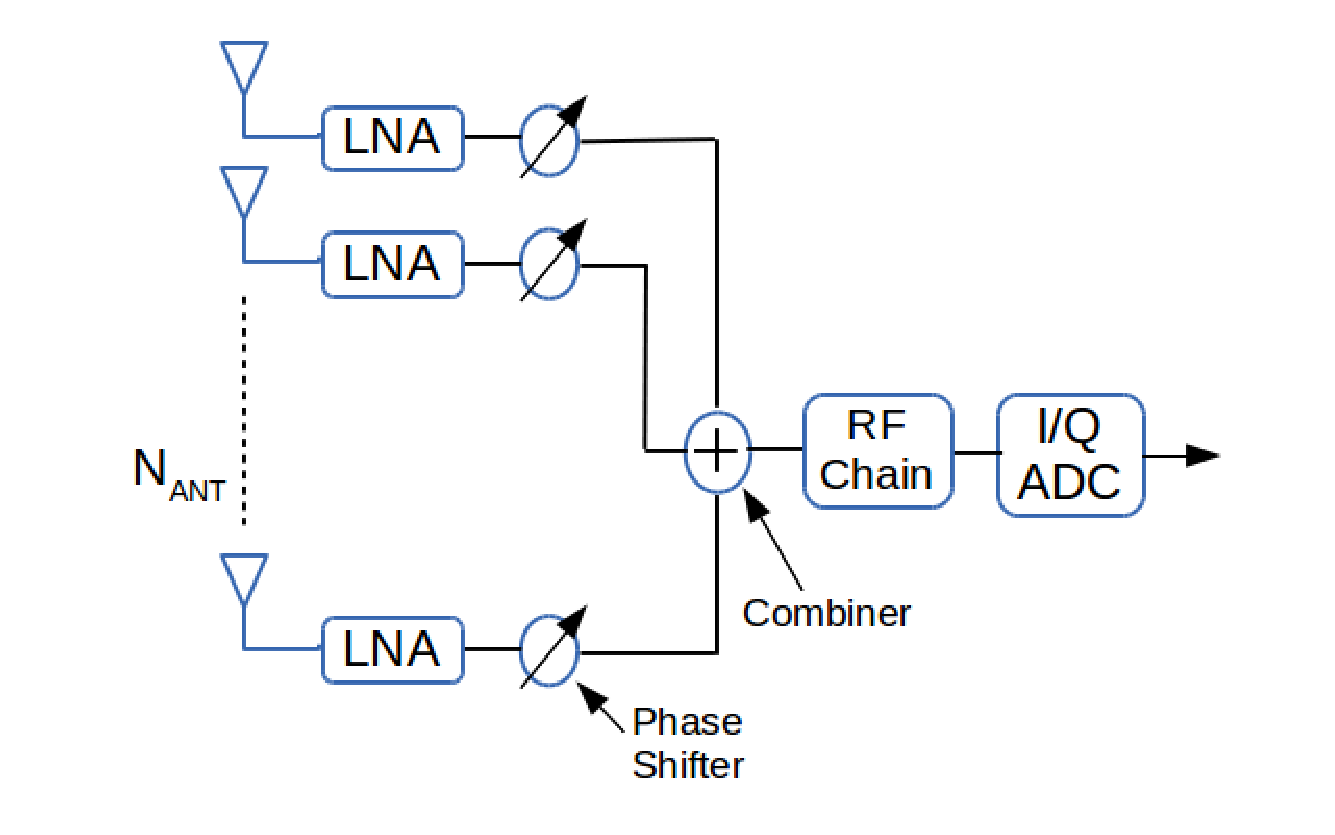}\vspace{-6mm}
        \caption{\protect\renewcommand{\baselinestretch}{1.25}\footnotesize  Analog Combiner.}\vspace{-4mm}
        \label{fig:AC}%
    \end{minipage}
    \hfill
    \begin{minipage}[t]{\wmpg}
    \psfrag{Nt = 64, Nr = 16, Npath = 8, HPADC}{\back \scriptsize $N_t = 64$, $N_r = 16$, $N_{paths} = 8$, HPADC}
        \includegraphics[width=\columnwidth]{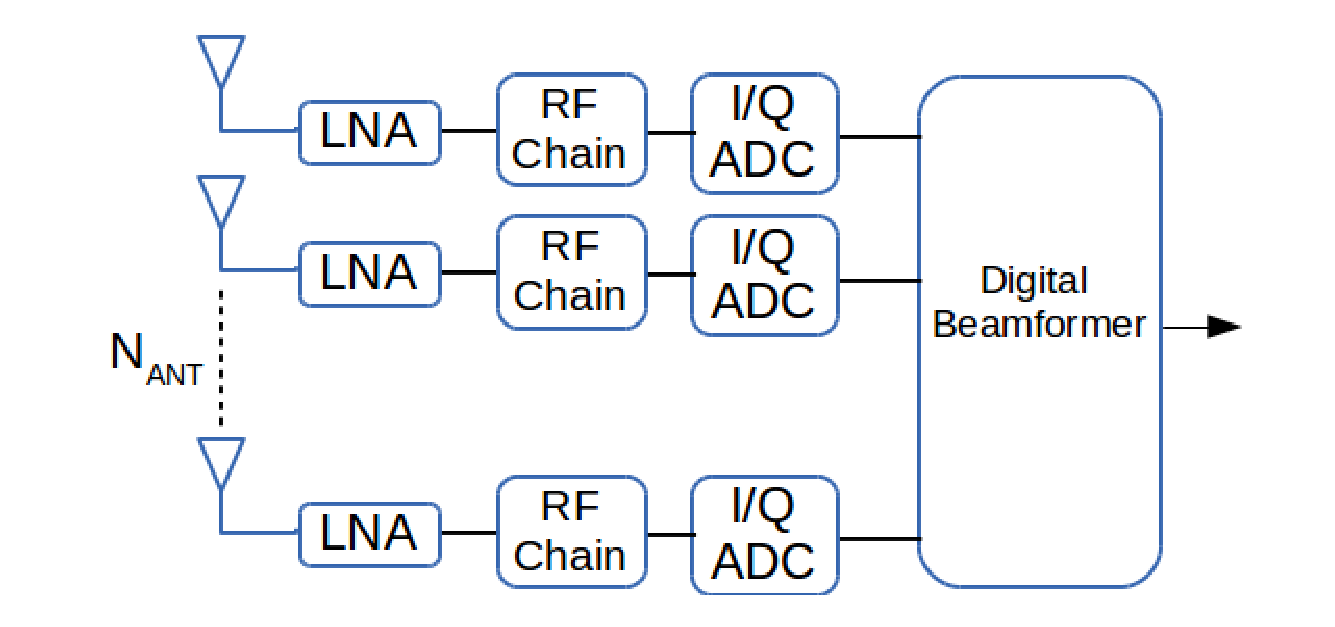}\vspace{-6mm}
        \caption{\protect\renewcommand{\baselinestretch}{1.25}\footnotesize Digital Combiner.}
        \label{fig:DC}%
    \end{minipage}\vspace{-5mm}
\end{figure}

\vspace{-8mm}

%
%
%
%

\begin{figure}
        \centering
        \scalebox{0.8}{\includegraphics[width=\columnwidth]{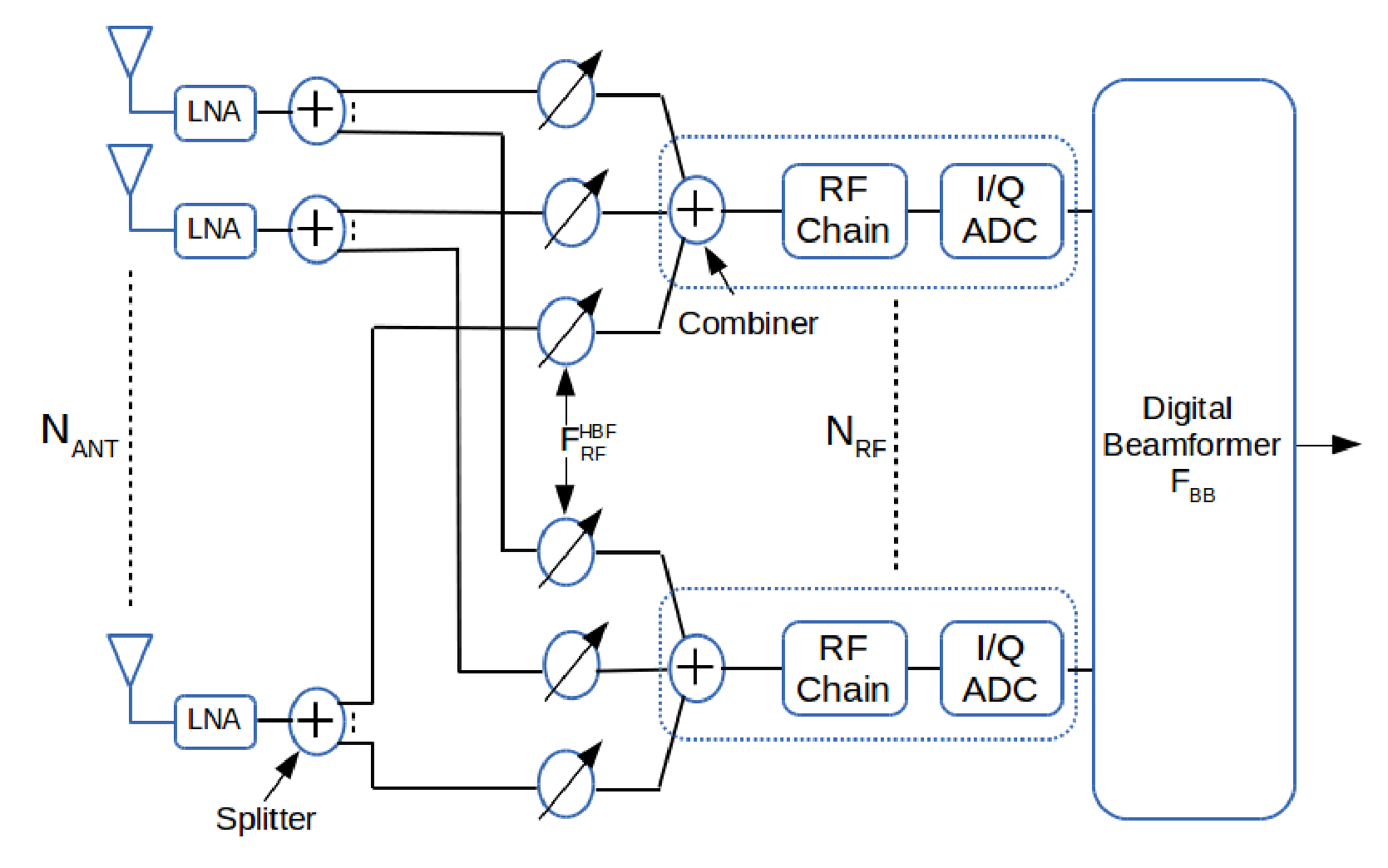}}\vspace{-4mm}
        \caption{\protect\renewcommand{\baselinestretch}{1.25}\footnotesize Hybrid Combiner.}
\label{fig:HC}\vspace{-5mm}
\end{figure}

\section{Receiver Architectures and Achievable Rates}
\label{sec:M_BF_C}

This section characterizes the achievable rates in mmW links with three types of receiver architectures featuring quantization. For all three cases, the transmitter architecture is considered to be always fully digital with ideal MIMO processing. We also consider the availability of the channel state information (CSI) both at the transmitter and at the receiver and we design the MIMO processing accordingly.

The three architectures are termed Analog, Digital, and Hybrid, represented in Figures \ref{fig:AC}, \ref{fig:DC} and \ref{fig:HC}, respectively. The difference between the three architectures consists in their different analog processing hardware prior to the ADC, which modifies the total number of RF and ADC units, and the number of digital signal dimensions that may be further processed by digital stages.

\subsection{Analog Combining}
\label{ssec:AC}

The first architecture is Analog Combining, which is motivated by the fact that typically ADCs are the most power-hungry receiver component. Therefore, in this architecture all multiple-antenna processing is performed in the analog domain to minimize power consumption. The architecture in Figure \ref{fig:AC} features one phase-shifter per receive antenna and an analog signal adder; together, these devices implement analog beamforming and deliver a scalar combined signal to a one-dimensional RF and ADC chain. 

The quantized received signal $\textbf{y}_q$ with analog combining at the receiver and digital beamforming at the transmitter is given by

\begin{equation}
{y}_q = (1-\eta)(\textbf{w}_r^H\textbf{H}\textbf{w}_t{x} + \textbf{w}_r^H\textbf{n}) + \textbf{n}_q
\label{eq:AC}
\end{equation} 

where superscript $H$ denotes conjugate transpose, $\textbf{w}_t$ represents the digital beamforming vector at the transmitter such that $||\textbf{w}_t||^2 = 1$ and $\textbf{w}_r$ is the analog combining vector at the receiver with a constant amplitude per coefficient $|w_{r}^i| = 1/\sqrt{N_r}$ due to its implementation using phase-shifters.

The additive quantization noise variance is given by $\eta(1-\eta)(|\textbf{w}_r^H\textbf{H}\textbf{w}_t|^2P + N_o)$, where $P$ is the average transmitter power. Finally, the ergodic capacity maximization problem with analog combining is given as

\begin{equation}
\begin{aligned}
	C_{AC}  = \mathrm{E}_{\mathbf{H}}\bigg[ \underset{\textbf{w}_{r},\textbf{w}_t}  {\max}\ &B\log_2\bigg(1+\dfrac{(1-\eta)|\textbf{w}_r^H\textbf{H}\textbf{w}_t|^2P}{{BN_o} + \eta|\textbf{w}_r^H\textbf{H}\textbf{w}_t|^2P}\bigg) \bigg]\\
	 s.t.\ &|w_{r,i}| = \frac{1}{\sqrt{N_r}}, \\
		   &||\textbf{w}_t||^2 = 1,	     
\end{aligned}
	\label{eq:Cq_AC}
\end{equation}

Due to the fact that CSI is available for each channel realization, and the fact that \eqref{eq:Cq_AC} is a monotonic function of the parameter $|\textbf{w}_r^H\textbf{H}\textbf{w}_t|^2$, the maximization of ergodic capacity is achieved by maximizing this beamforming gain independently for each channel realization. 

The transmitter beamforming vector, which has fewer constraints, can be simply assigned the value that maximizes the gain for a given value of $\mathbf{w}_r$. This consists in implementing a matched filter at the transmitter with value $\textbf{w}_t = \dfrac{\textbf{H}^H\textbf{w}_r}{||\textbf{H}^H\textbf{w}_r||^2}$, where the normalization is required to satisfy the transmit power constraint.

By using the matched filter at transmission, we can rewrite the problem as finding the receive beamforming vector that maximizes $|\textbf{w}_r^H\textbf{H}|^2$. If $\textbf{w}_r$ had no per-coefficient amplitude constraints, the optimal receive beamforming would be the eigenvector $\textbf{u}_{\max}$ associated with the largest singular value $\sigma_{\max}$ of the matrix $\textbf{H}$, i.e.,

\begin{equation}
\label{eq:optABFfree}
 \textbf{u}_{\max}=\arg\max_{\textbf{w}_r\in \mathbb{C}^{N_r}} |\textbf{w}_r^H\textbf{H}|^2,
\end{equation}.

However, since the analog scheme can only alter the phase of a constant-amplitude beamforming vector, the exact optimal analog beamforming vector is, in turn, expressed as

\begin{equation}
\label{eq:optABFphase}
 \textbf{w}_r^*=\arg\max_{\textbf{w}_r: |w_{r}^i| = 1/\sqrt{N_r}} |\textbf{w}_r^H\textbf{H}|^2.
\end{equation}

Finally, due to the fact that the problem in \eqref{eq:optABFphase} is more difficult, in our model we consider that the receiver instead settles for an approximate solution consisting in the projection of the unconstrained optimal beamforming vector $\textbf{u}$ from \eqref{eq:optABFfree} to the nearest point over the space of constant-amplitude vectors, i.e.,

\begin{equation}
\label{eq:optABFvec}
 \tilde{\textbf{w}}_r^H=\frac{1}{\sqrt{N_r}}(e^{\measuredangle u_{\max}^1},e^{\measuredangle u_{\max}^2}, \dots e^{\measuredangle u_{\max}^{N_r}})^T
\end{equation}


\subsection{Digital Combining}
\label{ssec:DC}

The second architecture is Digital Combining, which is motivated by the fact that digital MIMO processing has in general fewer constraints and can achieve higher gains. Therefore, in this architecture all multiple-antenna processing is performed in the digital domain to maximize the rate. The architecture in Figure \ref{fig:DC} features no analog processing; each antenna directly delivers its received signal to a dedicated RF and ADC chain. A quantized signal with $N_r$ dimensions is processed by a Digital MIMO processor that allows the spatial multiplexing of up to $N_s\leq \min(N_t,N_r)$ symbols streams. The quantized received signal with digital combining is given as

\begin{equation}
\textbf{y}_q = (1-\eta)(\textbf{W}_r^H\textbf{H}\textbf{W}_t\textbf{x} + \textbf{W}_r^H\textbf{n}) + \textbf{W}_r^H\textbf{n}_q
\label{eq:DC}
\end{equation}

where $\textbf{W}_r$ and $\textbf{W}_t$ are the digital combining and beamforming matrices, respectively. 
Note that, as combining is performed after quantization, the DC matrix also multiplies the quantization noise.
To calculate the supremum achievable rate with DC, we design the beamforming and combining matrices corresponding to the singular value decomposition of the channel matrix i.e., $\textbf{H} = \textbf{U}\Sigma\textbf{V}^H$, where $\textbf{U}$ and $\textbf{V}$ are the left and right singular matrices, respectively, and $\Sigma$ is a diagonal matrix with the singular values.
Now, by applying a transmit beamforming matrix $\textbf{W}_t = \textbf{V}$ and receive combining matrix $\textbf{W}_r = \textbf{U}^H$, Eq. (\ref{eq:DC}) can be written as

\begin{equation}
\begin{aligned}
\textbf{y}_q &= (1-\eta)(\textbf{U}^H\textbf{U}\Sigma \textbf{V}^H\textbf{V}\textbf{x} + \textbf{U}^H\textbf{n}) + \textbf{U}^H\textbf{n}_q \\
&= (1-\eta)(\Sigma\textbf{x} + \textbf{U}^H\textbf{n}) + \textbf{U}^H\textbf{n}_q
\end{aligned}
\label{eq:DC_SVD}
\end{equation}

Finally, we allocate the transmit power across the singular values of $\Sigma$ using the water filling algorithm, and the capacity with DC results in

\begin{equation}
	C_{DC}  = \mathrm{E}_{\mathbf{H}}\bigg[ \max_{\mathbf{Q}} B\log_2\det\bigg|\textbf{I}+\dfrac{(1-\eta)\Sigma\textbf{Q}\Sigma^H}{{BN_o\textbf{I}} + \eta\textbf{U}^H\diag(\textbf{U}\Sigma\textbf{Q}\Sigma^H\textbf{U}^H)\textbf{U}}\bigg|\bigg]	 
	\label{eq:Cq_DC}
\end{equation}


where due to the fact that $\textbf{Q}$ appears also in the noise, the $\textbf{Q}$ that maximizes this expression is \textit{not} the diagonal matrix containing the power levels computed with the water-filling algorithm for each channel realization. In our model, however, we employ the waterfilling algorithm, which may not be optimal, to obtain a lower bound approximation on the rates achievable with Digital Combining.

At low SNR, the waterfilling algorithm allocates all the power to the maximum singular value, and therefore the beamforming and combining vectors are just the right and left singular vectors ($\textbf{v}_{\max}$ and $\textbf{u}_{\max}$) corresponding to the maximum singular value, respectively. This shows how the optimal number of spatial streams $N_s$ may be smaller than $\min(N_t,N_r)$; compared with AC, which only provides beamforming gain, DC can provide the advantages of both spatial multiplexing gains at high SNR and beamforming power gains at low SNR. Although DC achieves a capacity similar to AC at low SNR, the former is always greater than the latter, due to the fact that the DC architecture does not impose constant-amplitude constraints in the receive combining vector. The capacity with DC as $SNR\to 0$ is given as

\begin{equation}
	C_{DC}  \stackrel{SNR\to 0}{=} \mathrm{E}_{\mathbf{H}}\bigg[B\log_2\bigg(1+\dfrac{(1-\eta)|\textbf{u}_1^H\textbf{H}\textbf{v}_1|^2P}{BN_o + \eta|\textbf{u}_1^H\textbf{H}\textbf{v}_1|^2P}\bigg) \bigg]
	\label{eq:Cq_DC_LSNR}
\end{equation}

\subsection{Hybrid Combining}
\label{ssec:HC}

The third architecture is Hybrid Combining, which is motivated by the high capacity and the low power consumption of Digital and Analog combining architectures, respectively, and tries to strike a balance between the two. Therefore, this architecture has both analog and digital multiple-antenna processing. The architecture in Figure \ref{fig:HC} features an analog processing stage with multiple banks of phase-shifters, each with an independent analog adder, RF and ADC chain. The analog processing reduces the dimensions of the received signal to a number $N_{RF}$ greater than one (analog case) but smaller than $N_{r}$ (digital case). The processed received signal with $N_{RF}$ dimensions is quantized and processed by a Digital MIMO processor that allows the spatial multiplexing of up to $N_s\leq \min(N_t,N_{RF})\leq N_{r}$ symbols streams.
The quantized signal with HC is given by

\begin{equation}
\textbf{y}_q = (1-\eta)(\textbf{F}_{BB}^H\textbf{W}_{RF}^H\textbf{H}\textbf{W}_t\textbf{x} + \textbf{F}_{BB}^H\textbf{W}_{RF}^H\textbf{n}) + \textbf{F}_{BB}^H\textbf{n}_q
\label{eq:HC}
\end{equation}  

where $\textbf{W}_{RF}$ and $\textbf{F}_{BB}$ are the RF and the baseband combining vectors, respectively.
Let $\textbf{n}_e = \textbf{W}_{RF}^H\textbf{n}$ represents equivalent receiver noise, $\textbf{H}_e = \textbf{W}_{RF}^H\textbf{H}$ represent a $N_{RF} \times N_t$ equivalent channel matrix, and $\textbf{H}_e = \textbf{U}\Sigma\textbf{V}^H$ represent its singular value decomposition.
The digital baseband combiner may be set to $\textbf{F}_{BB} = \textbf{U}^H$, and its corresponding transmit beamformer to $\textbf{W}_t = \textbf{V}$, which leaves the quantized signal in (\ref{eq:HC}) written as  

\begin{equation}
\textbf{y}_q = (1-\eta)(\Sigma\textbf{x} + \textbf{U}^H\textbf{n}_e) + \textbf{U}^H\textbf{n}_q
\label{eq:HC_SVD}
\end{equation}

Similar to the analog case, the design of $\mathbf{W}_{RF}$ is affected by the phase-shifter with constant amplitude constraints. The unconstrained optimization should maximize the result of a water-filling over the $N_{RF}$ singular values of the equivalent channel $\textbf{H}_e=\mathbf{W}_{RF}\mathbf{H}$. This results in the following optimization problem

\begin{equation}
\label{eq:optHBFfree}
 \mathbf{W}_{RF}^{ideal}=\arg\max_{\textbf{W}_{RF}\in \mathbb{C}^{N_r\times N_{RF}}} \max_{\sum p_i=P}  \sum_{\mathcal{S}(\mathbf{W}_{RF}\mathbf{H})} \sigma_i^2p_i,
\end{equation}

where $\mathcal{S}(\mathbf{W}_{RF}\mathbf{H})$ denotes the spectrum of the matrix and $\sigma_i^2$ is the ith eigenvalue. However, the analog processors can only alter the phase of constant-amplitude values of $\textbf{W}_{RF}$. This, in turn, is expressed as

\begin{equation}
\label{eq:optHBFphase}
 \textbf{W}_r^*=\arg\max_{\textbf{W}_{RF}: |W_{r}^{ij}| = 1/\sqrt{N_r}} \max_{\sum p_i=P} \sum_{\mathcal{S}(\mathbf{W}_{RF}\mathbf{H})} \sigma_i^2p_i.
\end{equation}

Since the latter is more difficult to compute, our model considers a practical approximate solution of \eqref{eq:optHBFphase} based on the projection of the $N_{RF}$ eigenvectors associated with the $N_{RF}$ largest singular values of $\mathbf{H}$, $\textbf{u}_1\dots \textbf{u}_{N_{RF}}$ over the space of constant-amplitude matrices. Notice that, differently from the analog case, this projection does not necessarily have to correspond to the phase of the $N_{RF}$ columns of the solution to \eqref{eq:optHBFfree}. The equivalent RF combining matrix $\tilde{\textbf{W}}_{RF}$ is given as

\begin{equation}
\label{eq:optHBFmat}
 \tilde{\textbf{W}}_{RF}=\frac{1}{\sqrt{N_r}}\left(\begin{array}{ccc}
 e^{\measuredangle u_1^1}& \dots &e^{\measuredangle u_1^{N_r}}\\
 \vdots& \ddots& \vdots\\
 e^{\measuredangle u_{N_{RF}}^1}& \dots & e^{\measuredangle u_{N_{RF}}^{N_r}}\\
 \end{array}\right)^T
\end{equation}

This results in a similar $\textbf{y}_q$ as obtained with DC in (\ref{eq:DC_SVD}).
However, $\textbf{U}$ has a dimension of $N_{RF}$ instead of $N_r$ (as was the case in DC), due to the limited number of RF chains. 
The capacity with HC is given as  

\begin{equation}
	C_{HC}  = \mathrm{E}_{\mathbf{H}}\bigg[B\log_2\det\bigg|\textbf{I}+\dfrac{(1-\eta)\Sigma\textbf{Q}\Sigma^H}{{BN_o\textbf{I}} + \eta\textbf{U}^H\diag(\textbf{U}\Sigma\textbf{Q}\Sigma^H\textbf{U}^H)\textbf{U}}\bigg|\bigg]	 
	\label{eq:Cq_HC}
\end{equation}

Although the two capacity expressions for DC and HC are similar, the capacity of HC is upper bounded by the capacity of DC due to the fact that the selection of $\textbf{W}_{RF}$ is subject to a constant amplitude constraint associated with the analog combiners.

\subsection{Capacity Results}
\label{ssec:CapRes}
\begin{figure}[t]
    \centering
    \begin{minipage}[t]{\wmpg}
    \psfrag{Nt = 64, Nr = 64, Nc = 1, Np = 1}{\back \scriptsize $N_t = 64$, $N_r = 64$, $N_{c} = 1$, $N_{p} = 1$}
        \includegraphics[width=\columnwidth]{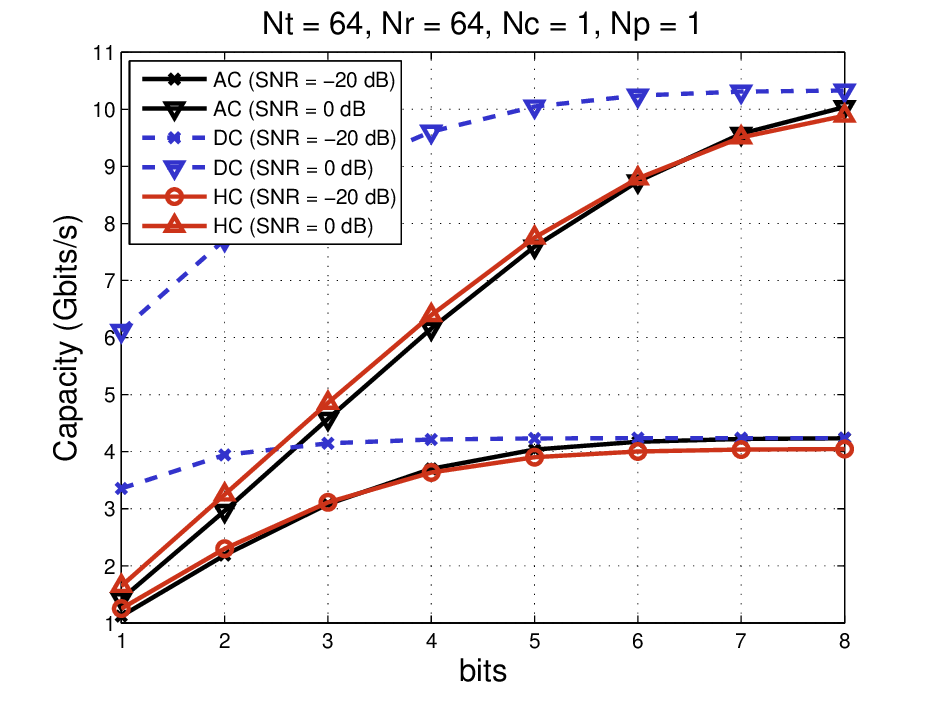}\vspace{-6mm}
        \caption{\protect\renewcommand{\baselinestretch}{1.25}\footnotesize  Capacity vs ADC bits comparison for AC, DC and HC schemes for $N_p =1$.}%
        \label{fig:CapNpath1}%
    \end{minipage}
    \hfill
    \begin{minipage}[t]{\wmpg}
    \psfrag{Nt = 64, Nr = 64, Nc =1, Np = 10}{\back \scriptsize $N_t = 64$, $N_r = 64$, $N_{c} = 1$, $N_{p} = 10$}
        \includegraphics[width=\columnwidth]{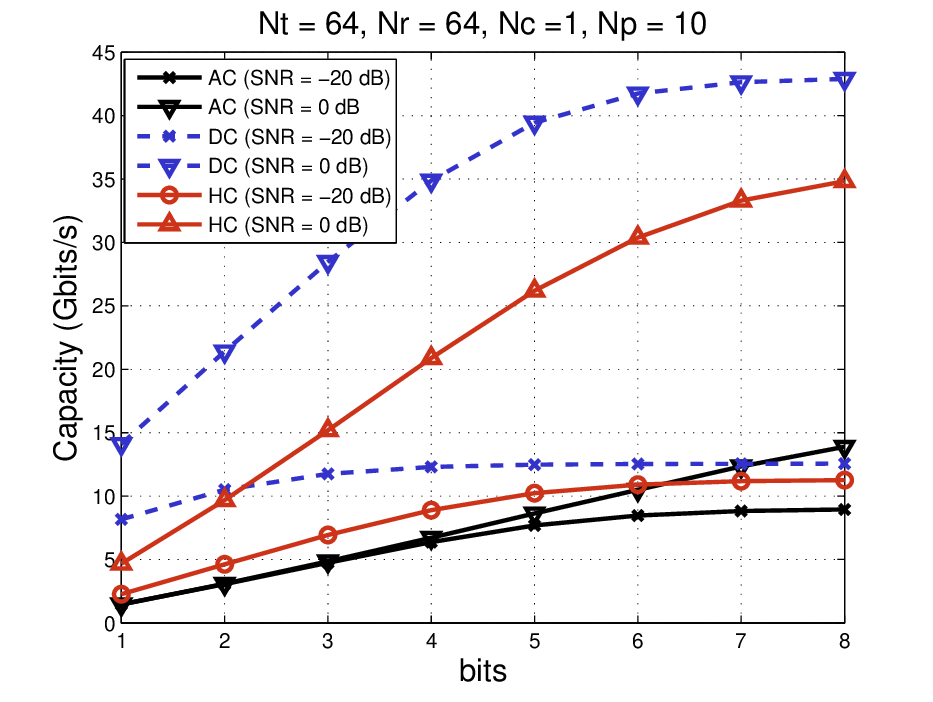}\vspace{-6mm}
        \caption{\protect\renewcommand{\baselinestretch}{1.25}\footnotesize  Capacity vs ADC bits comparison for AC, DC and HC schemes for $N_p =10$.}
        \label{fig:CapNpath8}%
    \end{minipage}\vspace{-5mm}
\end{figure}

We compare the average achievable rates for AC, DC and HC mmW links as a function of the number of ADC bits.
Figures \ref{fig:CapNpath1} and \ref{fig:CapNpath8} show the rate vs ADC bits when the number of propagation paths is $N_p = 1$ and $N_p = 10$, respectively.
In this simulation we consider $N_t = 64$, $N_r = 64$, $N_{RF} = 4$, $N_c = 1$, Bandwidth $B = 1$ GHz and the results are averaged over 1000 independent channel realizations. The total transmit power is set to 30 dBm and we show results for $-$20 and 0 dB SNR\footnote{Note that $-$20 and 0 dB SNR does not include antenna gain and therefore corresponds to an approximate communication range of 100 m for NLOS and LOS, respectively.}. 

In both figures we can see that, for all architectures, the capacity grows up to a certain number of ADC bits and saturates afterwards, so that a further increase in $b$ does not improve the SNR of the quantized signal. The saturation threshold appears later when the unquantized signal suffers from weaker noise, and the number of bits required to reach this saturation increases with the SNR of the unquantized signal, $\gamma$.  

The results show that the DC architecture outperforms the other schemes for any number of propagation paths. This is due to the fact that DC does not have constant amplitude constraints in the beamforming coefficients. Moreover, in the case of $N_p = 1$, i.e., the channel has rank $1$, spatial multiplexing is not possible, and AC can perform similarly to HC or DC, but only for a high number of ADC bits. In the opposite case, for $N_p = 10$, AC achieves significantly lower rates because it cannot exploit spatial multiplexing, unlike HC and DC. This shows that the appeal of AC schemes is strongest for single-path or sparse propagation environments.
Finally, even though HC exploits spatial multiplexing, its rate is always slightly lower than DC due to the constant amplitude constraint of the RF combining matrix ($\textbf{W}_{RF}$) and also to the fact that DC allows spatial multiplexing of up to $\min(N_t,N_r)$ different streams.



\section{Energy Efficiency Analysis}
\label{sec:EE}

Looking only at capacity, the straightforward choice for a mmW receiver design would be a fully digital architecture (i.e., DC), which can exploit the maximum advantages of both beamforming and spatial multiplexing techniques, outperforming AC and HC. However, generally, these advantages of DC are tied to a higher power consumption at the receiver. Thus, although DC results in the maximum achievable rates, it may not be an energy efficient receiver option. 

In particular, in the large bandwidth operation expected in mmW, the ADC is usually considered to be the most power hungry block and thus the power consumption of DC is penalized by its high number of ADCs ($N_{ADC}$), equal to twice the number of receive antennas. In comparison, AC, only requiring $N_{ADC} = 2$, would be the the least power consuming scheme, and HC, requiring $N_{ADC} = 2\times N_{RF}$, is generally assumed to have a power consumption in-between AC and DC.

Nonetheless, looking only at the ADC, and disregarding the power consumption of analog entities not necessary for DC, may be misguided.  Particularly, if the power consumption of phase shifters and analog combiners is non-negligible, HC may be heavily penalized due to the fact that it requires a large number (up to $N_{RF}$) of both analog blocks and ADCs at the same time.

In this section we study the EE  of the three architectures, defined as
\begin{equation}
   EE =  \dfrac{C_q}{P_{tot}}
   \label{eq:EE}
\end{equation} 

where $C_q$ is the capacity of the quantized signal corresponding to different combining schemes and $P_{Tot}$ is the total power consumption of the mmW receiver design corresponding to analog, digital and hybrid combining architectures.

\begin{table}
\centering
    \begin{minipage}[t]{\wmpg}
\caption{Power consumption of each device}
\label{tab:devicepowers}
\begin{tabular}{rll}
Device & Notation & Sim. Value\\ \hline
Low Noise Amplifier (LNA)\cite{Rx_Pow_LNA_PS_C} & $P_{LNA}$ & $39$mW\\
Splitter & $P_{SP}$ & $19.5$mW\\
Combiner \cite{Rx_Pow_LNA_PS_C}& $P_{C}$ & $19.5$mW\\
Phase shifter \cite{Rx_Pow_LNA_PS_C}& $P_{PS}$ & $19.5$mW\\
Mixer \cite{Rx_Pow_60GHz}& $P_{M}$ & $16.8$mW\\
Local oscillator \cite{RialandHeath} & $P_{LO}$ & $5$mW\\
Low pass filter \cite{RialandHeath}& $P_{LPF}$ & $14$mW\\
Base-band amplifier \cite{RialandHeath}& $P_{BB_{amp}}$ & $5$mW\\
ADC & $P_{ADC}$ & $cB2^{b}
$\\
\end{tabular}
    \end{minipage}
    \hfill
    \begin{minipage}[t]{\wmpg}
 \centering
 \caption{ADC power per sample and per level}
 \label{tab:adcparam}
 \begin{tabular}{rll}
 Scenario & Value & Generation\\ \hline
 LPADC & $5$fJ/step/Hz & Ideal future value\\
 HPADC & $494$fJ/step/Hz & State of the art\\
 PHPADC& $12.5$pJ/step/Hz & Last decade legacy\\
 \end{tabular}
    \end{minipage}\vspace{-5mm}
\end{table}

\subsection{Power Consumption Model}
The devices required to implement each mmW receiver architecture are displayed in Figures \ref{fig:AC}, \ref{fig:DC} and \ref{fig:HC}, respectively.
The total power consumption $P_{Tot}$ of each scheme is evaluated by the following expressions
\begin{equation}
   P_{Tot}^{AC} =  N_{r}(P_{LNA}+P_{PS}) + P_{RF} + P_C+ 2P_{ADC}
   \label{eq:P_AC}
\end{equation}  
\begin{equation}
\begin{split}
   P_{Tot}^{HC} = &\ N_{r}(P_{LNA} + P_{SP} + N_{RF}P_{PS})\\			& + N_{RF}(P_{RF}+P_C + 2P_{ADC} )
\end{split}
\label{eq:P_HC}   
\end{equation}
\begin{equation}
   P_{Tot}^{DC} =  N_{r}(P_{LNA} + P_{RF} + 2P_{ADC})
   \label{eq:P_DC}
\end{equation} 
\normalsize
where $P_{RF}$ represents the power consumption of one RF chain, given by
\begin{equation}
   P_{RF} =  P_M + P_{LO} + P_{LPF} + P_{BB_{amp}},
   \label{eq:Prf}
\end{equation}
\normalsize
and the component power consumptions are detailed in Table \ref{tab:devicepowers}. The power consumption of all components except the ADC is independent of the bandwidth $B$ and the number of bits $b$, whereas
$P_{ADC}$ increases exponentially with the number of bits $b$ and linearly with the bandwidth $B$ and with
the ADC Walden's figure of merit $c$ \cite{ADC_b_B} (the energy consumption per conversion step per Hz). 

In our analysis we shall consider three different generations of ADC technology with varying Walden's figure, detailed in Table \ref{tab:adcparam}. The Low Power ADC (LPADC) model considers the estimated lowest limit to the power needs of ADCs at $1$ GHz, deduced by the hardware survey \cite{ADCs_97-15} and cited by analyses such as \cite{Ahmed16_EE}; the High Power ADC (HPADC) model is based in a modern device that supports sampling of Gs/s and has been referenced in recent mmW literature, considered in \cite{OrhanER15PowerCons}; and finally the Pessimistic HPADC (PHPADC) model assumes the use of a decade-old ADC design, prior to recent advancements in technology. PHPADC represents a worst-case scenario for digital architecture, but is close to the ADC power consumption model used in some literature about HC. Due to the pessimistic nature of this power model, we believe such literature may have overestimated the advantages of HC. For a comprehensive survey on ADC hardware the reader is referred to \cite{ADCs_97-15}.  


\begin{figure}[t]
    \centering
    \begin{minipage}[t]{\wmpg}
    \psfrag{Nt = 64, Nr = 64, Nc = 1, Np = 10, HPADC}{\back \back \ \ \scriptsize $N_r = 64$, $N_{RF} = 4$, $N_{c} = 1$, $N_{p} = 10$, HPADC}
        \includegraphics[width=\columnwidth]{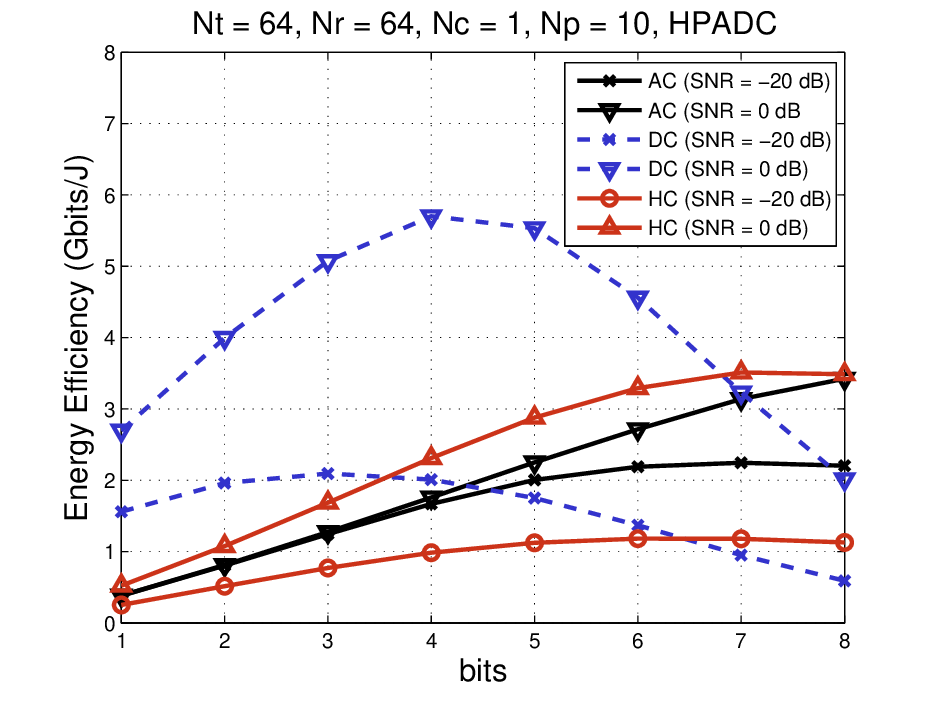}\vspace{-6mm}
        \caption{\protect\renewcommand{\baselinestretch}{1.25}\footnotesize  EE vs.ADC bits comparison for AC, DC and HC schemes for a HPADC model.}%
        \label{fig:EE_HPADC}%
    \end{minipage}
    \hfill
    \begin{minipage}[t]{\wmpg}
    \psfrag{Nt = 64, Nr = 64, Nc =1, Np = 10, PHPADC}{\back\back \ \ \scriptsize $N_r = 64$, $N_{RF} = 4$, $N_{c} = 1$, $N_{p} = 10$, PHPADC}
        \includegraphics[width=\columnwidth]{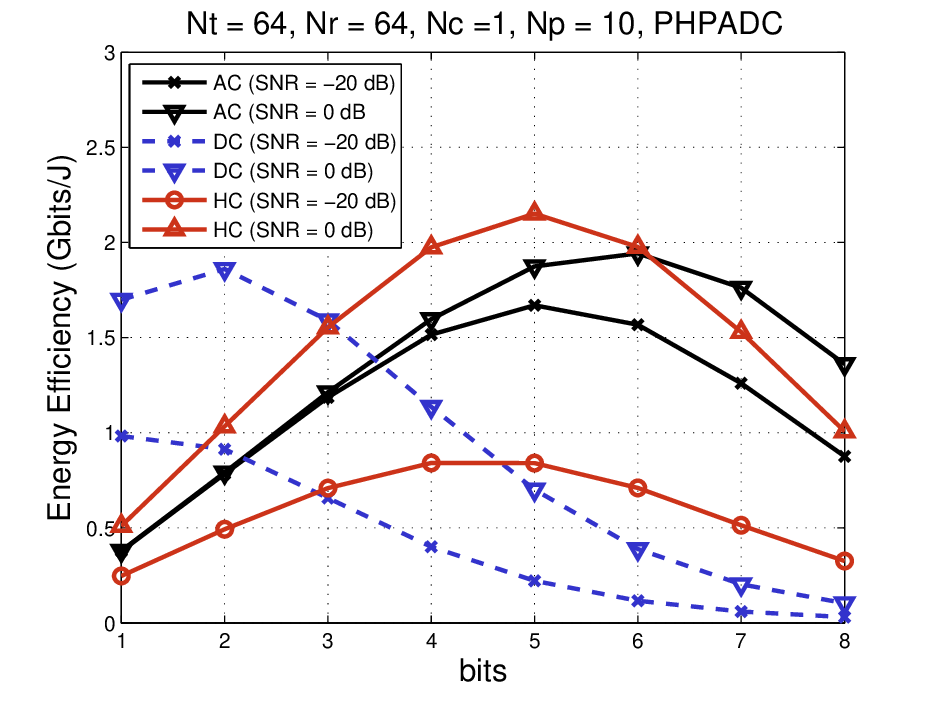}\vspace{-6mm}
        \caption{\protect\renewcommand{\baselinestretch}{1.25}\footnotesize EE vs.ADC bits comparison for AC, DC and HC schemes for a PHPADC model.}
        \label{fig:EE_PHPADC}%
    \end{minipage}\vspace{-5mm}
\end{figure}

\subsection{EE Results}
\label{ssec:EE_Res}

We analyze the EE of each receiver architecture using the power values per component defined in Tables \ref{tab:devicepowers} and \ref{tab:adcparam}. We show the EE vs number of ADC bits for HPADC and PHPADC scenarios in Figures \ref{fig:EE_HPADC} and \ref{fig:EE_PHPADC}, for two SNR values, $-$20 dB and 0 dB, respectively. In this analysis we have considered a mmW antenna array with $N_t = 64$, and $N_r = 64$, a hybrid scheme with $N_{RF} = 4$, a channel with bandwidth $B = 1$ GHz and a number of propagation paths set to $N_p = 10$, and $N_c =1$. 

The plots show that all combining schemes have an optimal number of ADC bits which results in a maximum EE, with EE increasing with $b$ before this value, but decreasing for any further increase in $b$ above the optimal. The optimal point is influenced by both the saturation of capacity as $b$ increases, which depends on SNR, and by the ADC power consumption model, that increases with $b$.

Comparing the two ADC generations, we note that the legacy PHPADC model power consumption increases at a higher rate with $b$ than the HPADC model, while capacity remains the same irrespective of the ADC power model. Therefore, the optimal number of bits for any scheme must be smaller using higher power-consuming ADCs.

Comparing the SNR, we note that, everything else being equal, the capacity saturation effect occurs for a higher number of bits when the SNR is larger, and therefore the optimal number of bits for highest EE is increased with SNR too. 
  
Finally, comparing the three receiver architectures, we note that surprisingly DC offers the highest EE of the three schemes with a HPADC model (Figure \ref{fig:EE_HPADC}) whereas results in a similar 
EE to HC and AC with PHPADC (Figure \ref{fig:EE_PHPADC}), but its peaks occur for very small optimum number of bits. By increasing $b$ beyond the optimal for DC, the EE of the fully digital archiecture decays very rapidly, and comparing protocols with the same number of bits, when $b$ is not very low, DC is outperformed by AC and HC in terms of EE because DC requires more ADCs and therefore its power consumption grows more rapidly with $b$.

Moreover, when the SNR is low, AC has a better performance than HC, due to the fact that at low SNR the water-filling algorithm focuses all the power in the single strongest singular value of the channel, limiting the spatial multiplexing advantage of HC. At the same time, HC uses more analog components while its capacity is similar to AC, and thus the EE may be worse.

It must be noted that when DC offers the highest EE, with fewer bits, its capacity is lower than the ceiling value the same architecture achieves when $b$ grows. For this reason, in the next section we focus on the simultaneous comparison of the protocols in the two dimensions (SE and EE), using two dimensional charts to characterize the regions of operation of each architecture.

\section{Simulation Results}

In this section we develop a comparison  of DC, HC, and AC receiver architectures in two dimensions at the same time. To this end, we create comparison charts that represent the EE and the SE  of each receiver design, allowing to study the choice of the appropriate combining scheme, and to observe how its performance depends on the different system parameters. 

In the chart, for each architecture, we plot curves representing the evolution of its SE versus EE performance as we increase the number of ADC bits $b$, from 1 to 8, at increments of 1. The highest points in the chart correspond to highest capacity or SE. The rightmost points in the chart correspond to highest EE. Thus, the closer a point to the top-right corner, the better the corresponding protocol; and the bigger the area contained at the left of the curve, the more the versatility of an architecture as a function of the number of bits. In addition, an absolute power constraint may be displayed over the chart as a straight line, where only points below the line satisfy the power constraint.

\subsection{AC is superior only if $N_c=1$}

We will first study the effect of the number of independent paths between the transmitter and the receiver, seeing that AC is only superior when propagation is rank one. 

\begin{figure}[t]
    \centering
    \begin{minipage}[t]{\wmpg}
    \psfrag{Nt = 64, Nr = 64, Nc =1, Np = 1, PHPADC}{\back\back \ \ \scriptsize $N_r = 64$, $N_{RF} = 4$, $N_{c} = 1$, $N_{p} = 1$, PHPADC}
        \includegraphics[width=\columnwidth]{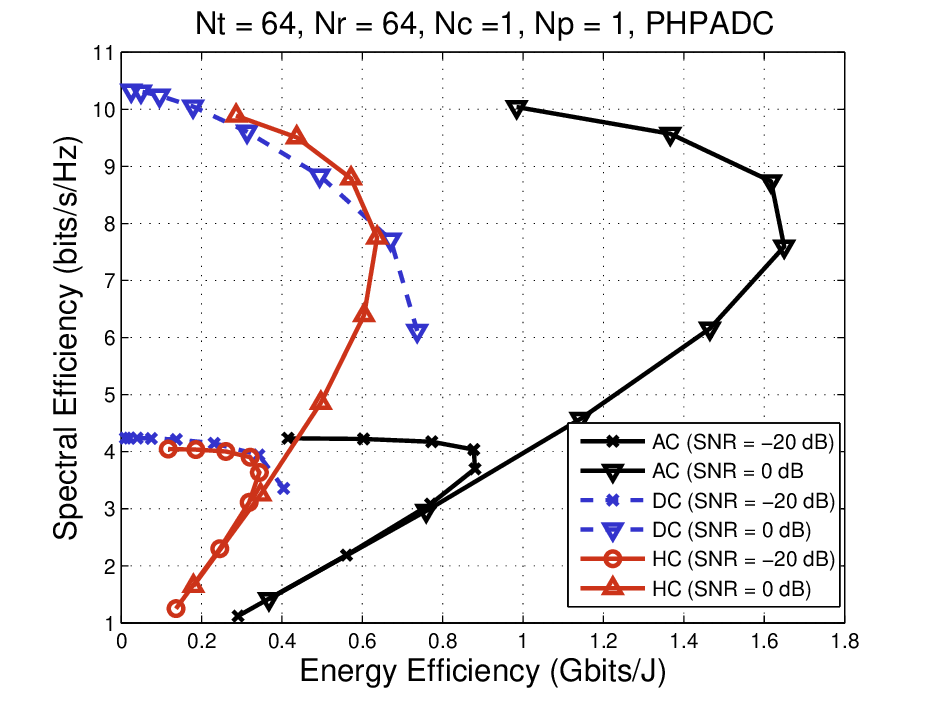}\vspace{-6mm}
        \caption{\protect\renewcommand{\baselinestretch}{1.25}\footnotesize SE vs. EE comparison for AC, DC and HC schemes for a PHPADC model with $N_{p} = 1$ and $N_{RF} = 4$ for HC.}%
        \label{fig:EEvsSE_Npath1}%
    \end{minipage}
    \hfill
    \begin{minipage}[t]{\wmpg}
    \psfrag{Nt = 64, Nr = 64, Nc =1, Np = 10, PHPADC}{\back\back \ \scriptsize $N_r = 64$, $N_{RF} = 4$, $N_{c} = 1$, $N_{p} = 10$, PHPADC}
        \includegraphics[width=\columnwidth]{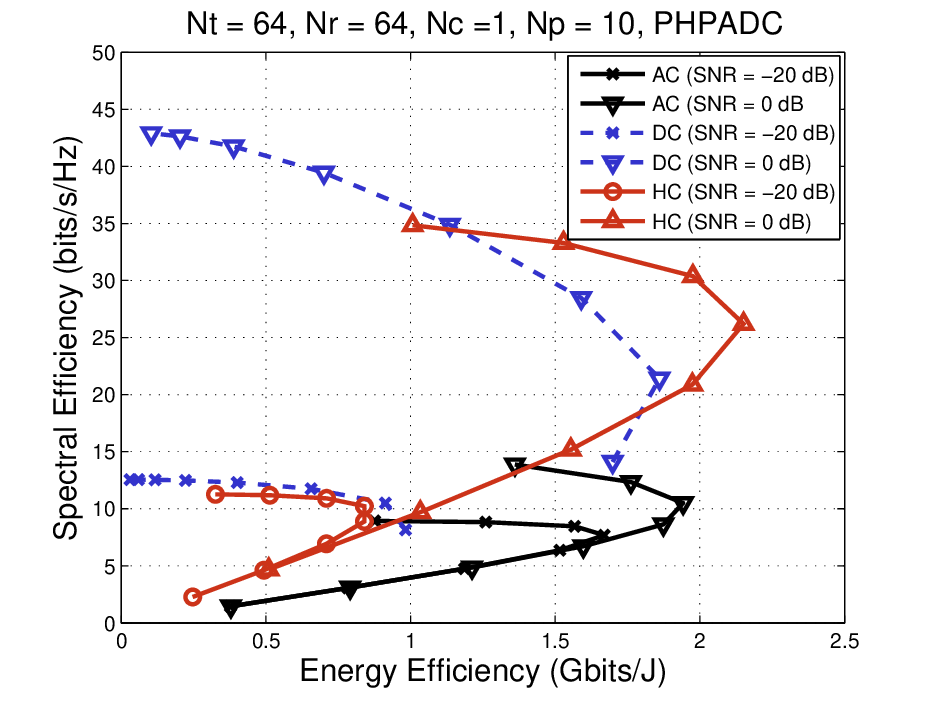}\vspace{-6mm}
        \caption{\protect\renewcommand{\baselinestretch}{1.25}\footnotesize SE vs. EE comparison for AC, DC and HC schemes for a PHPADC model with $N_{p} = 10$ and $N_{RF} = 4$ for HC.}
        \label{fig:Worst_DC}%
    \end{minipage}\vspace{-5mm}
\end{figure}

Figures \ref{fig:EEvsSE_Npath1} and \ref{fig:Worst_DC} compare SE vs EE charts for $N_p = 1$ and $N_p = 10$, respectively, using the PHPADC model, a single cluster, and $N_{RF} =4$ for HC. In the charts we observe curves that evolve as the number of bits $b$ is increased, reaching upward and right up to a point where they wrap around themselves and EE returns to the top left corner. Each curve can be interpreted as an ``operating region'' where receivers of each design architecture can be implemented. The more extended the area covered by a receiver, the more versatile the corresponding design; and the closer a curve to the top-right corner, the better the corresponding architecture.

We observe that EE only increases up to a certain number of bits due to the fact that at small $b$, capacity increases more than the power, and EE increases. However,  at higher $b$ power consumption increases faster than the capacity, resulting in a decrease of the EE while the SE still increases. This occurs due to the fact that the power consumption of an ADC increases exponentially whereas the SE increases sub-linearly with an increase in the number of ADC bits.
  
We note that for $N_p = 1$ the spectral efficiencies reached by the three architectures are fairly similar, but AC displays much better energy efficiency at both SNR = $-$20 dB and SNR = 0 dB. The AC curves cover a much larger area, reaching points with similar rates and much better energy efficiencies than HC and DC for any number of bits.
This occurs due to the fact that at $N_p=1$ the channel has only one singular value, there is no spatial multiplexing gain, capacities are fairly similar, and the receiver with the least hardware tends to be the better. However, this phenomenon is exclusive to the rank-1 channel, and disappears when there is a spatial-multiplexing gain. As Figure \ref{fig:Worst_DC} shows, the performance of AC with respect to DC and HC starts degrading with an increase in $N_p$.  

Finally, note that there is only a small region on the graphs where HC performs better than DC (i.e., HC covers a region not reached by DC) for both $N_p = 1$ and $N_p =10$. This is due to two effects:  1) the power consumption of DC is comparable or even less than HC when the number of bits is low, and 2) the SE of HC is slightly lower than DC due to the constant-amplitude constraints of analog processing. In the following comparisons, we develop a systematic characterization of all the scenarios where DC outperforms HC.

\subsection{HC versus DC: A Baseline}
\label{sec:baseline}

Our results show that a sparse propagation with one path is the  only scenario where AC comes close to the performance of DC and HC. For the rest of the tests, we shall focus on the differences between DC and HC. For this, we define a \textit{baseline} that is the most favorable case for HC architectures, where indeed HC outperforms DC in a specific sense, and we will show that the advantage of HC is very fragile and in all use-cases that deviate from this baseline DC becomes better. Figure \ref{fig:Worst_DC} shows this reference case with a pessimistic ADC power consumption (PHPADC), a very large number of receive antennas $N_{r} = 64$ that is not matched by a small number of propagation paths $N_p = 10$ and a single cluster (i.e., $N_c = 1$); and with a small number of HC RF chains, i.e., $N_{RF} = 4$, which is a better match with channel spatial dimensions.


Figure \ref{fig:Worst_DC} shows the result for this scenario for both SNR = $-$20 dB and SNR = 0 dB. In the chart, the larger area covered and the closer to the top-right corner, the better. The result shows that HC beats DC at high SNR when we take both into account, reaching closer to the top-right corner, thus achieving a much larger EE while the SE is nearly as good. Particularly, HC can achieve the most efficient point with SE $=27$ bps/Hz and EE $=2.2$ Gb/J, or increase the rate to 33 bps/Hz while penalizing EE to 1.5 Gb/J. In comparison, DC at the same 35 bps/Hz achieves only 1 Gb/J (requiring 50\% more power for the same rate) and DC with the maximum number of bits, despite offering 43 Gbps/Hz, has a very low energy efficiency (0.2 Gb/J, 1000\% more power that the most efficient scheme).

Note that if we made a point to point comparison with the same number of bits, up to $b = 3$, specific points with DC would achieve better values than the equivalent points of HC. However, the full two-dimensional chart proves its value as a comprehensive representation by the fact that we can clearly see that there are other values of $b$ where HC can reach into the same region and outperform all points of DC. This example shows how architecture comparison must not be performed point-vs-point on the same number of bits, but region-vs-region over the entire chart.

AC is always the scheme displaying the least power consumption in absolute terms (bottom right corner), but its SE is lower and it is only a preferable scheme for a low power system design where reaching a higher rate is less important than extending the battery life.

Finally, the chart also shows that even with all conditions stacked against a digital scheme, there are particular instances when DC may still be a preferable option than HC. For instance, at low SNR, where only few bits are enough to achieve rates similar to those of an unquantized model, the point for 2 bits in DC (0.92,10) achieves $\approx$ 10\% more EE than HC with 5 bits (0.84,10). Another example at high SNR is the small triangular region below HC and above AC, only achievable with the DC architecture with 1 bit.

In the following comparisons, we show that these instances where DC is superior become the dominant case once any  parameter deviates from the baseline. This means that the advantage of HC is very fragile, and if even just one of the mmW system characteristics drifts away from the scenario where the odds are purposefully stacked in favor of HC, then DC becomes the clearly superior receiver architecture.

\subsection{Improvement of ADCs}

For the second comparison, we study the influence of the ADC power consumption generations. We compare the pessimistic, decade-old ADC power model in the baseline from Section \ref{sec:baseline} with a more current state of the art and a future prediction ADC technology model; that is, we compare LPADC versus HPADC and PHPADC, and show that even for just a moderate improvement in ADC technology, DC clearly  outperforms HC.

\begin{figure}[t]
    \centering
    \begin{minipage}[t]{\wmpg}
    \psfrag{Nt = 64, Nr = 64, Nc =1, Np = 10, HPADC}{\back\back \ \ \scriptsize $N_t = 64$, $N_r = 64$, $N_{c} = 1$, $N_{p} = 10$, HPADC}
        \includegraphics[width=\columnwidth]{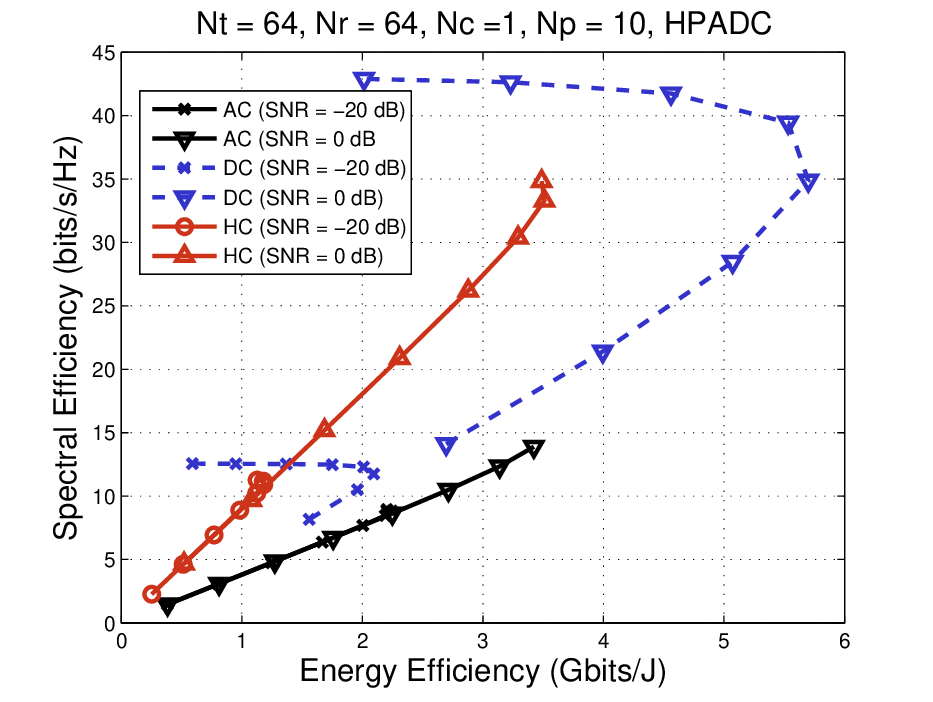}\vspace{-6mm}
        \caption{\protect\renewcommand{\baselinestretch}{1.25}\footnotesize  SE vs EE comparison for AC, DC and HC schemes for a HPADC model.}%
        \label{fig:EEvsSEHPADC}%
    \end{minipage}
    \hfill
    \begin{minipage}[t]{\wmpg}
    \psfrag{Nt = 64, Nr = 64, Nc =1, Np = 10, LPADC}{\back\back \ \ \scriptsize $N_t = 64$, $N_r = 64$, $N_{c} = 1$, $N_{p} = 10$, LPADC}
        \includegraphics[width=\columnwidth]{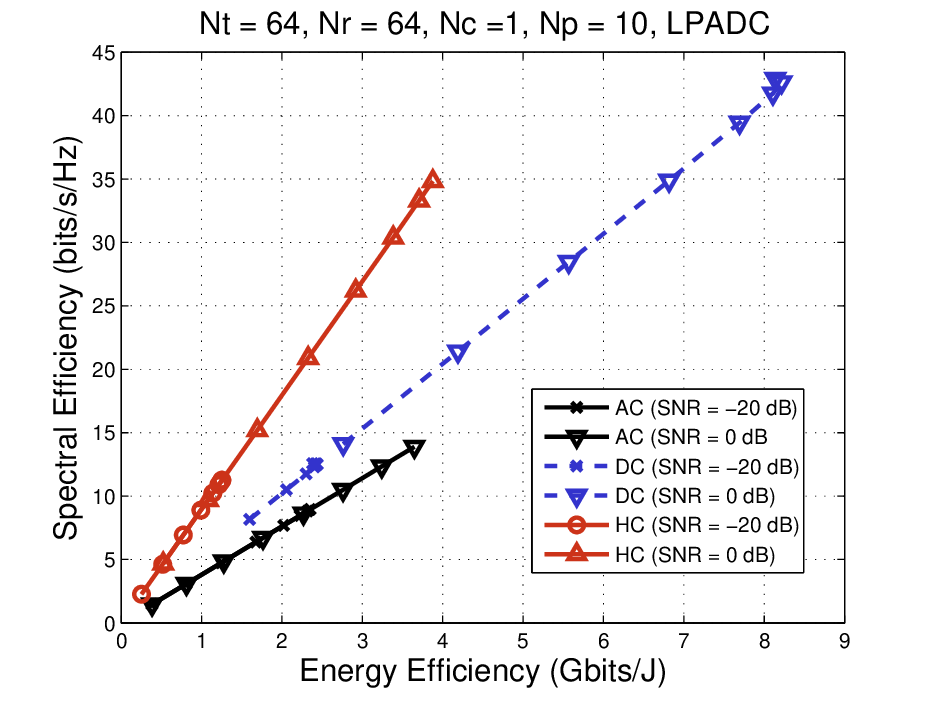}\vspace{-6mm}
        \caption{\protect\renewcommand{\baselinestretch}{1.25}\footnotesize SE vs EE comparison for AC, DC and HC schemes for a LPADC model.}
        \label{fig:EEvsSELPADC}%
    \end{minipage}\vspace{-5mm}
\end{figure}

Figures \ref{fig:EEvsSEHPADC} and \ref{fig:EEvsSELPADC} show the SE vs EE plot for a HPADC and a LPADC models. The charts 
show that, in general, the EE improves with the decrease in ADC power consumption for any combining scheme, although the improvement is more relevant to DC, which starts dominating the HC scheme for all levels of SE and EE. Note that the entire EE range greatly improves with the ADC figure of merit, from 0-2 in the baseline to 0-5 in HDADC and 0-9 in LPADC. This is exploited mostly by DC, which reaches the most top-right regions of the chart.

In a point-to-point comparison with the same number of bits, for 7 or 8 bits and under the HPADC model, the EE of DC (3.2 and 2 Gb/J with 43 bps/Hz respectively) is lower than HC (3.5 Gb/J at 35 bps/Hz), but again, the chart shows that there exist points where DC reaches much better EE with similar SE and fewer bits (4.5 Gb/J at 42 bps/Hz and 6 bits, for instance). Again, the big picture comparison region-vs-regions shows the differences more accurately.

Finally, with a LPADC model where the power consumption of ADC is even less significant than the power consumption of other components, DC always provides a better EE and SE than other combining schemes, even for the same number of bits $b$. For LPADC, the ADCs power consumption is so low that the curves do not wrap, and it is possible to increase the number of bits up to $8$ while increasing both EE and SE at the same time.

Here it must be remarked that the baselines for the rest of our analysis below only focus on the PHPADC, as we have clearly shown that if ADCs improve, DC is superior. In the next comparisons, we show that even if ADC power is pessimistic, DC can outperform HC when the parameters $N_r$, $N_{RF}$ and $N_c$ are varied.

\begin{figure}
        \centering
        \psfrag{Nt = 64, Nr = 16, Nc =1, Np = 10, PHPADC}{\ \ \ \ \  $N_r = 16$, $N_{RF} = 4$, $N_{c} = 1$, $N_{p} = 10$, PHPADC}
        \scalebox{0.7}{\includegraphics[width=\columnwidth]{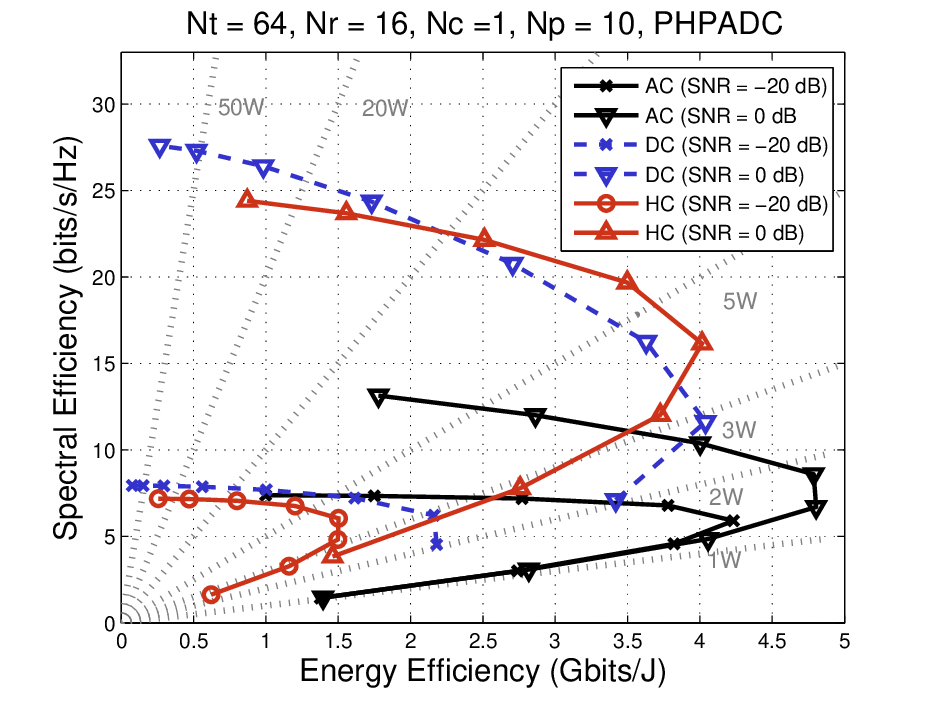}}\vspace{-4mm}
        \caption{\protect\renewcommand{\baselinestretch}{1.25}\footnotesize SE vs EE for a PHPADC model with $N_r = 16$, $N_{RF} = 4$.}
\label{fig:EEvsSENr16} \vspace{-5mm}
\end{figure}

\subsection{Reduced number of receive antennas}

In the third comparison we study the influence of the number of receive antennas. Due to the fact that the propagation has few paths, it may be wasteful to use a full 64 antenna array at the receiver. Indeed, if the receiver has only 16 antennas (or turns on only the first 16 of a larger array), DC outperforms HC completely in low-SNR systems and achieves greater rates in high-SNR. The use of a fixed subset of 16 antennas in a 64 array may be regarded as a further simplification of antenna selection techniques, an alternative to HC proposed in \cite{RialandHeath}. 

Figure \ref{fig:EEvsSENr16} shows SE vs EE trend for AC, DC and HC schemes for $N_r = 16$, $N_{RF} = 4$ and $N_p =10$ while considering a PHPADC model.
Results show that with a decrease in $N_r$ SE decreases while EE increases (as compared to $N_r = 64$ in Figure \ref{fig:Worst_DC}) for all combining schemes.
This is because a reduction in $N_r$ reduces both the array gain and the power consumption of any combining scheme.
The reduction in array gain reduces the SE logarithmically while a decrease in the power consumption (which is linear with $N_r$) increases the EE.

In comparison to Figure \ref{fig:Worst_DC}, results also show that with a decrease in $N_r$, the difference between the maximum EE for HC and DC decreases significantly. Now, at high SNR, the difference between HC and DC has narrowed down. DC can achieve 4 Gb/J at 12.5 bps/Hz with 2 bits, whereas HC offers the same 4 Gb/J at 16 bps/Hz with 4 bits. This, in turn, is close to the point with 16 bps/Hz and 3.7 Gb/J for DC. For low SNR, DC completely contains the operational region of HC and for any number of bits of HC a point with better EE or SE or both can be found on the DC curve. 
For a reduced number of antennas, AC results in a supremum EE, with 4.2 Gb/J at low SNR and 4.8 Gb/J at high SNR. However, these points have very low SE (6 bps/Hz) due to their lack of exploitation of spatial multiplexing (unlike in the DC and HC schemes).

Figure \ref{fig:EEvsSENr16} shows how, in certain conditions, the decision between protocols can be difficult, as the comparison chart can be heterogeneous, displaying different operational regions that are covered by different architectures. We have added some constant-power rules to the chart to assess the different regions. Note that for a given power constraint, only points below the rule may be employed. First, below $2$ W, AC is the only viable architecture. Second, in the range $2-4$ W, there is a small region where DC outperforms HC. Third, from $4$ to $10$ W, HC outperforms DC. And finally, above $10$ W, DC outperforms HC again. Thus, surprisingly, DC is a better receiver for both smaller devices (such as UE up to 4 W) and larger devices (such as macro cell BS with 10, 20 or 50 W), while HC is better for medium power devices such as pico cell BS. If UEs cannot exceed 2 W, then AC should be used\footnote{Note that in this paper the energy efficiency is computed based on the power consumption values mentioned in Tables \ref{tab:devicepowers} and \ref{tab:adcparam}. However, to further investigate the achievable EE and to identify the appropriate operating regimes for different combining schemes, the reader is referred to our web viewing tool: https://dl.dropboxusercontent.com/u/1770302/mmWaveADCwebviewer/index.html} .

\subsection{Increased scattering reflections}
\label{ssec:EEvsSENc}
In the fourth comparison we challenge the baseline best scenario for HC with a scenario where the number of propagation paths, and rank of the channel, is larger than the number of HC RF chains. Therefore DC is able to increase the multiplexing gain beyond what is possible to HC, and again the fully digital architecture displays superiority.

\begin{figure}
        \centering
        \psfrag{Nt = 64, Nr = 64, Nc =2, Np = 10, PHPADC}{\ \ \ \ \ \  $N_r = 64$, $N_{RF} = 4$, $N_{c} = 2$, $N_{p} = 10$, PHPADC}
        \scalebox{0.7}{\includegraphics[width=\columnwidth]{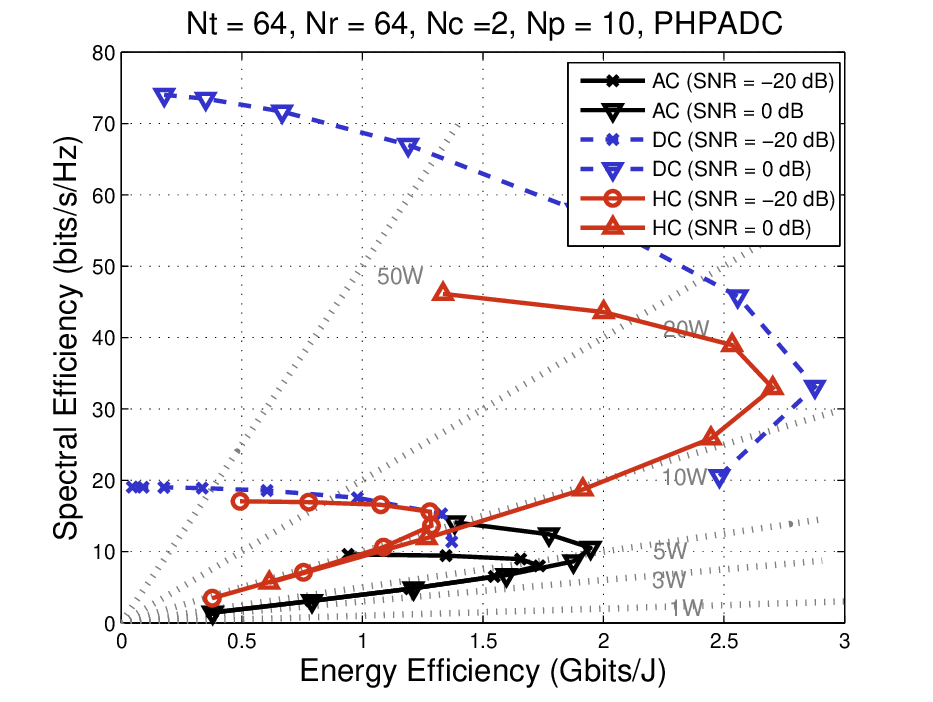}}\vspace{-4mm}
        \caption{\protect\renewcommand{\baselinestretch}{1.25}\footnotesize SE vs EE for a PHPADC model with $N_r = 64$, $N_{RF} = 4$, $N_{c} = 2$ and $N_{p} = 10$.}
\label{fig:EEvsSENpath8Nrf4}\vspace{-5mm}
\end{figure}

As an example, Figure \ref{fig:EEvsSENpath8Nrf4} shows the results with $N_p = 10$, $N_c =2$ and $N_{RF} =4$.

Observe that an increase in $N_c$ and $N_p$ also results in a significant increase in EE and SE, and that DC increases much more than HC. For instance at high SNR, HC now achieves 2.7 Gb/J at 35 bps/Hz with 7 bits in its most EE point, but DC achieves 2.85 Gb/J with the same rate and 2 bits, or 45 bps/Hz with an EE of 2.55 Gb/J and 3 bits.

This is because a fully digital architecture with $N_t = 64$ and $N_r = 64$ allows the spatial multiplexing of up to 64 symbol streams, and therefore an increase in $N_c$ also corresponds to an increase in the number of spatial multiplexing symbols, which increases SE for DC.
On the other hand, HC with only 4 RF chains allows spatial multiplexing of up to 4 symbols, and therefore when the channel matrix $\textbf{H}$ has more than $4$ dominant eigenvalues, HC cannot exploit them all and the improvement of its SE is marginal.
Moreover, an increase in channel rank only increases SE while the power consumption remains unchanged, and therefore EE also increases.
Finally, note that spatial multiplexing is more advantageous at high SNR, and therefore the increase in SE with an increase in $N_c$ for DC is higher at high SNR.

We see that, even for a mild increase in scattering diversity not matched by $N_{RF}$, DC becomes unquestionably superior to HC.

\subsection{Increased number of HC RF chains}

In the fifth and final comparison we modify the number of HC RF chains to match the increase in propagation paths featured in the previous comparison. 

\begin{figure}
        \centering
        \psfrag{Nt = 64, Nr = 64, Nc =2, Np = 10, PHPADC}{\ \ \ \ \ \  $N_r = 64$, $N_{RF} = 8$, $N_{c} = 2$, $N_{p} = 10$, PHPADC}
        \scalebox{0.7}{\includegraphics[width=\columnwidth]{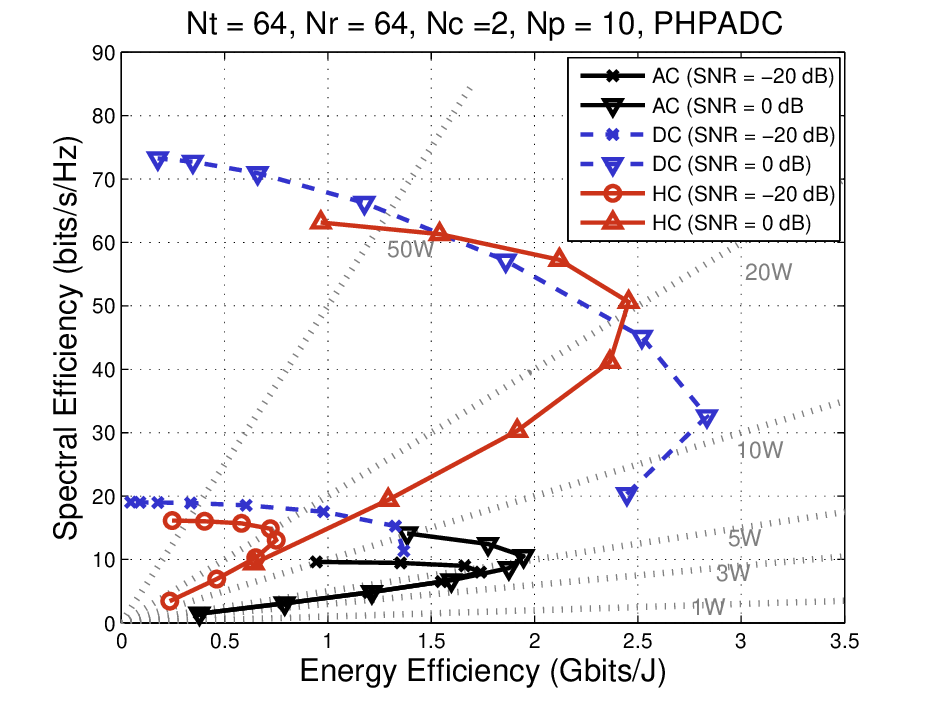}}\vspace{-5mm}
        \caption{\protect\renewcommand{\baselinestretch}{1.25}\footnotesize SE vs EE for a PHPADC model with $N_r = 64$, $N_{RF} = 8$, and $N_c = 2, N_p =10$.}
\label{fig:EEvsSENrf8Npath8}\vspace{-5mm}
\end{figure}

Figure \ref{fig:EEvsSENrf8Npath8} shows the results with $N_c = 2$, $N_p = 10$, and $N_{RF} = 8$. The curves of DC are identical due to the fact that power consumption of DC remains the same as in the previous section, whereas for HC, the number of analog phase-array blocks increases with $N_{RF}$, this increases the power consumption of HC. 

In comparison to Figure \ref{fig:EEvsSENpath8Nrf4}, the results show an improvement in SE for HC while EE decreases. For instance, at high SNR now HC achieves 50 bps/Hz with an EE under 2.5 Gb/J. An increase in SE directly corresponds to the availability of a higher number RF chains, which allows spatial multiplexing across more independent paths. However, this increase in $N_{RF}$ also increases the power consumption of HC scheme (Eq. \ref{eq:P_HC}) which eventually results in a similar or even slightly lower EE compared to what HC  achieved 
with $N_c = 2, N_p = 10, N_{RF} = 4$ (Figure \ref{fig:EEvsSENpath8Nrf4}).

As in the previous case, DC shows the best performance for both EE and SE (with 2 and 8 bits, achieving 35 bps/Hz at 2.85 Gb/J and 73 bps/Hz at 0.2 Gb/J, respectively). But now, at high SNR DC is not superior when both EE and SE are desired at the same time, as HC has a remarkably interesting small region not contained in the curve of DC near the top-right corner, corresponding to about 50 bps/Hz and 2.5 Gb/J. At low SNR, DC always outperforms HC.

Adding constant-power rules, we can observe that at high SNR HC should be chosen for $20-50$ W (macro BS), DC for $10$-$20$ W (pico BS), and only AC can operate below $5$ W. At low SNR, DC and AC meet in the 10 W rule too.

It is important to compare the last two figures at once. We have that, when the number of propagation paths of the channel grows, there are two options for HC, namely; a) Maintaining $N_{RF}$, the EE is good but SE is low, whereas b) Increasing $N_{RF}$, the SE is good but the EE is penalized. Neither option allows HC to maintain its advantage over DC, and thus the digital architecture works better when the scattering is not-so-poor. This is intuitive since, in the limit as $N_c=N_{RF}=N_{r}$, a HC scheme is just a wasteful implementation of DC with unnecessary hardware.
 
\section{Conclusions}
\label{sec:concl}
In this work, we studied the spectral and energy efficiency trade-off for analog, digital and hybrid combining schemes.
The results show that there is only one scenario where AC is better, and only one scenario where HC is better, whereas for any other mmW channel and hardware scenario, DC outperforms HC in several aspects. There is a surprisingly wide regime that favors the usage of DC with a low number of bits, which does not seem to be accurately represented by the popular claim that DC consumes the highest power, based only on ADC power consumption with a high number of bits. 

The ideal scenario for AC is a receiver with very tight power constraints in a mmW rank-1 channel. The former may be relevant for future low-power systems such as machine-type communications, but the latter may not be entirely plausible, considering that with a $<10$ mm wavelength anything bigger than a human hand constitutes a potentially relevant reflector.

We have shown that the conventional wisdom that hybrid combining architectures are preferable over fully digital ones is, in fact, not so wise. The model for ADC power consumption varies significantly in mmW MIMO literature, and the ideal scenarios for HC in the literature have employed rather old available power efficiency figures of merit for ADCs, not considering the orders-of-magnitude improvements with current technology. When the ADCs improve even slightly, DC proves a much superior receiver architecture.

Even with pessimistic ADCs, we have also shown that if the receiver has a smaller antenna array, or if a subset of the antennas may be turned off to save power, the advantage of HC is, at best, not universal, and many cases can benefit from a DC architecture. Namely, low-SNR schemes, and devices with power constraints in the lowest and highest ends.

Moreover, even with pessimistic ADCs, we have shown that if the scattering environment is not-so-sparse (e.g., even in the presence of a mere 2-cluster channel matrix with 10 dependent reflections per cluster), the spatial multiplexing gains of DC achieve either a higher rate than HC (if the hybrid scheme uses fewer RF chains) or a similar rate with better energy efficiency (if the hybrid scheme increases the number of RF chains). Recall that the accurate model has an average of 1.9 clusters, with 3 clusters not being rare, and \textit{twenty} paths per cluster, and thus the levels of scattering diversity in our calculations are quite conservative in favor of HC.




\end{document}